\documentclass[aps,pre,preprint,groupedaddress]{revtex4-2}

\usepackage{graphicx}
\usepackage{bm}
\usepackage{amssymb}
\usepackage{amsbsy}
\usepackage{amsmath}
\usepackage{mathtools}
\usepackage{xcolor}
\usepackage{color}
\usepackage{tikz}
\usepackage[normalem]{ulem}
\usepackage{hyperref}

\makeatletter 
\gdef\@ptsize{1}
\let\@currsize\normalsize 
\makeatother 
\usepackage{setspace} 
\singlespace 

\begin{document}

\def\tr{\textcolor{red}}
\def\tb{\textcolor{blue}}


\title{The influence of active agent motility on SIRS epidemiological dynamics}


\author{R. Kailasham}
\thanks{Present address: Department of Chemical Engineering, Indian Institute of Technology Indore, Khandwa Road, Simrol, Madhya Pradesh 453552, India.}
\affiliation{Department of Chemical Engineering, Carnegie Mellon University, Pittsburgh, PA 15213, USA}
\author{Aditya S. Khair}
\email{akhair@andrew.cmu.edu}
\affiliation{Department of Chemical Engineering, Carnegie Mellon University, Pittsburgh, PA 15213, USA}



\date{\today}

\begin{abstract}
Active Brownian disks moving in two dimensions that exchange information about their internal state stochastically are chosen to model epidemic spread in a self-propelled population of agents under the susceptible-infected-recovered-susceptible (SIRS) framework. The state of infection of an agent, or disk, governs its self-propulsion speed; consequently, the activity of the agents in the system varies in time. Two different protocols (one-to-one and one-to-many) are considered for the transmission of disease from the infected to susceptible populations. The effectiveness of the two protocols are practically identical at high values of the infection transmission rate. The one-to-many protocol, however, outperforms the one-to-one protocol at lower values of the infection transmission rate.  Salient features of the macroscopic SIRS model are revisited, and compared to predictions from the agent-based model. Lastly, the motility induced phase separation in a population of such agents with a fluctuating fraction of active disks is found to be well-described by theories governing phase separation in a mixture of active and passive particles with a constant fraction of passive disks.
\end{abstract}

\maketitle

\section{\label{sec:intro} Introduction}

Epidemiological models have been in use for more than a century to track the spread of infectious diseases~\cite{Kermack1927}. These models assume that the population is compartmentalized into three populations: \textit{susceptible} (S) individuals, who are prone to the disease, \textit{infected} (I) individuals who possess the disease and are capable of spreading it, and individuals who have \textit{recovered} (R) from the disease and may or may not become prone to further infection~\cite{Anderson1991}. The rates of interconversion between the various populations are specified by epidemiological constants, and the time-evolution of these populations is governed by coupled ordinary differential equations (ODEs)~\cite{Anderson1991}. Several modifications to this susceptible-infected-recovered (SIR) model have been made, to account for policy interventions such as vaccination drives, social distancing protocols, and lockdown mandates~\cite{Cabrera2021}. SIR models and its variants have been used to describe the spread of several diseases~\cite{Bjrnstad2020}, such as: the plague~\cite{Kermack1927}, COVID-19~\cite{Wu2022,Liu2023}, varicella~\cite{Giraldo2008} and influenza~\cite{Osthus2017}. These macroscopic models assume a uniform, well-mixed population, such that the epidemiological constants do not depend on the (heterogeneous) spatial density of the various populations. A fine-grained, discrete treatment of epidemic spread is agent-based modeling~\cite{Peruani2008,Peruani2019,Gonzalez2004,Rodriguez2019} (ABM), which considers each individual in the population as an autonomous entity, with the disease transmission occurring if a susceptible individual comes within a contagion radius of the infected individual. {Another paradigm involves the use of network-based modeling to account for spatial heterogeneities in the population and variabilities in the epidemiological constants~\cite{Kiss2017,Grossmann2021,7393962,Pastor-Satorras2015}.} More recently, SIR modeling has been coupled~\cite{Forgacs2022,Zhao2022,Norambuena2020} with advances in the understanding of self-propelled or \textit{active matter}. Self-propelled entities such as light-activated Janus particles undergo a transition from a homogeneous state to a clustered one, at sufficiently high enough values of packing fraction and self-propulsion speed~\cite{Buttinoni2013,Bialke2015}. This experimentally observed phenomenon of motility induced phase separation (MIPS) can be recapitulated in numerical simulations in which the active particles translate with a constant self-propulsion speed, interact sterically, and whose orientations evolve diffusively in time~\cite{Cates2015,Zottl2023}. It is therefore possible, using such systems, to examine disease propagation in both homogeneous and spatially heterogeneous configurations, by attaching to the individual active agents an internal state (S, I or R).
If the internal state of the particle is coupled to its dynamics, such that the self-propulsion speed of infected particles is different from that of susceptible or recovered particles, then one would expect that MIPS in such a system would be affected. We address this question in the present work, and also aim to draw a connection between the agent-based, or microscopic, and macroscopic descriptions of epidemic spread.

\begin{figure}[t]
\begin{center}
\begin{tabular}{c c}
\includegraphics[width=3in,height=!]{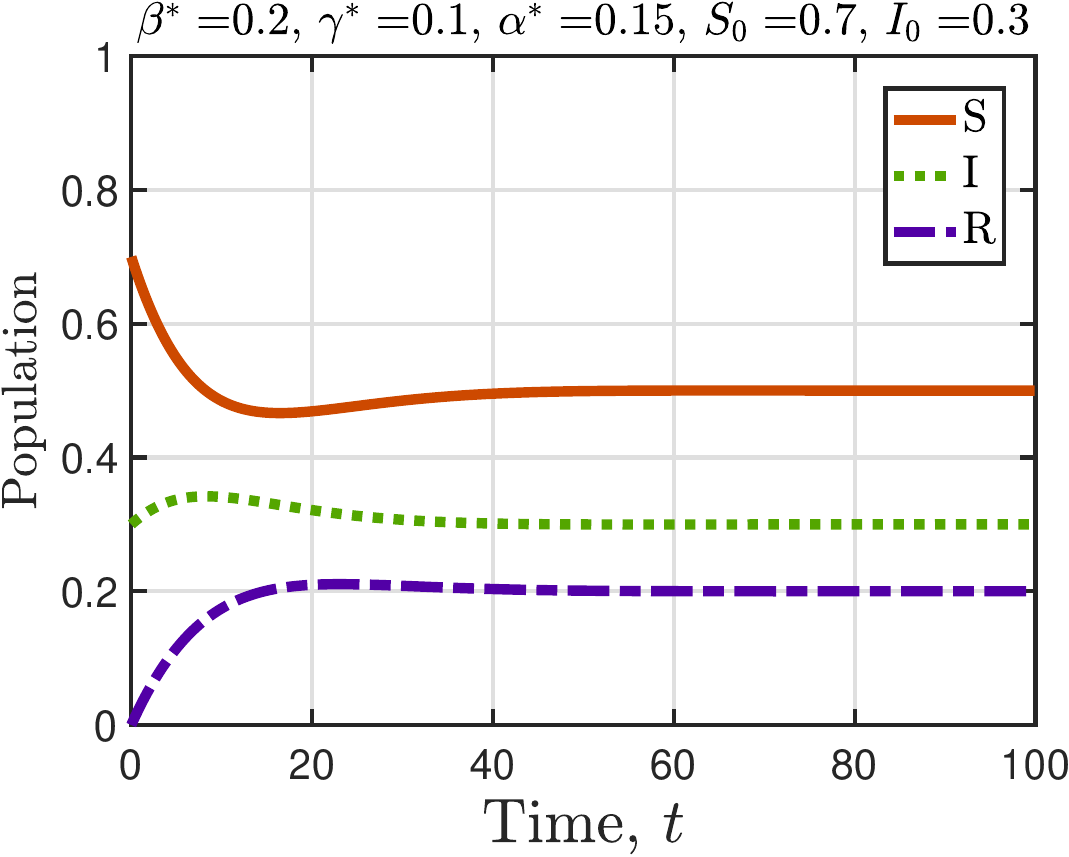}&
\includegraphics[width=3in,height=!]{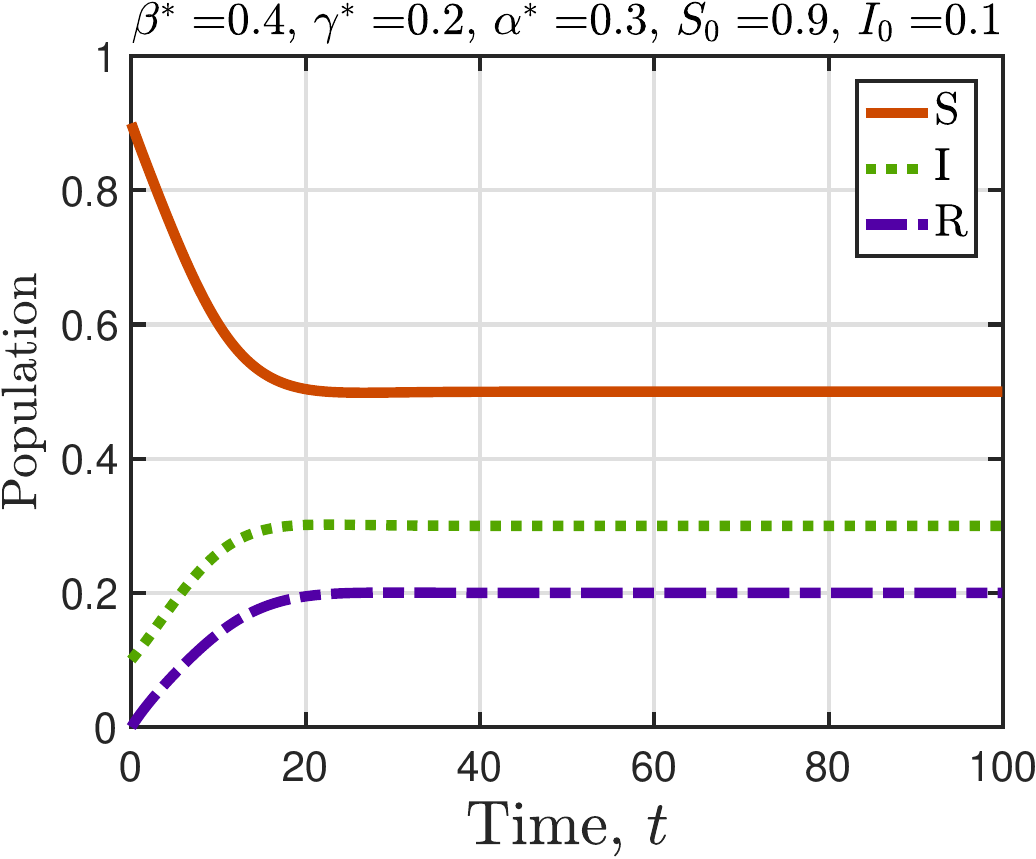}\\
(a) & (b) \\[5pt]
\end{tabular}
\end{center}
\caption{Population evolution as a function of time for two parameter sets with $\omega^{*}=2$, $\mu^{*}=1.5$ and different values of the initial populations of susceptible and infected individuals, i.e., (a) $S_{0}=0.7$, $I_{0}=0.3$ and (b) $S_{0}=0.9$, $I_{0}=0.1$.}
\label{fig:tevol_macro}
\end{figure}

\begin{figure}[t]
\begin{center}
\begin{tabular}{c c c}
\includegraphics[width=2.1in,height=!]{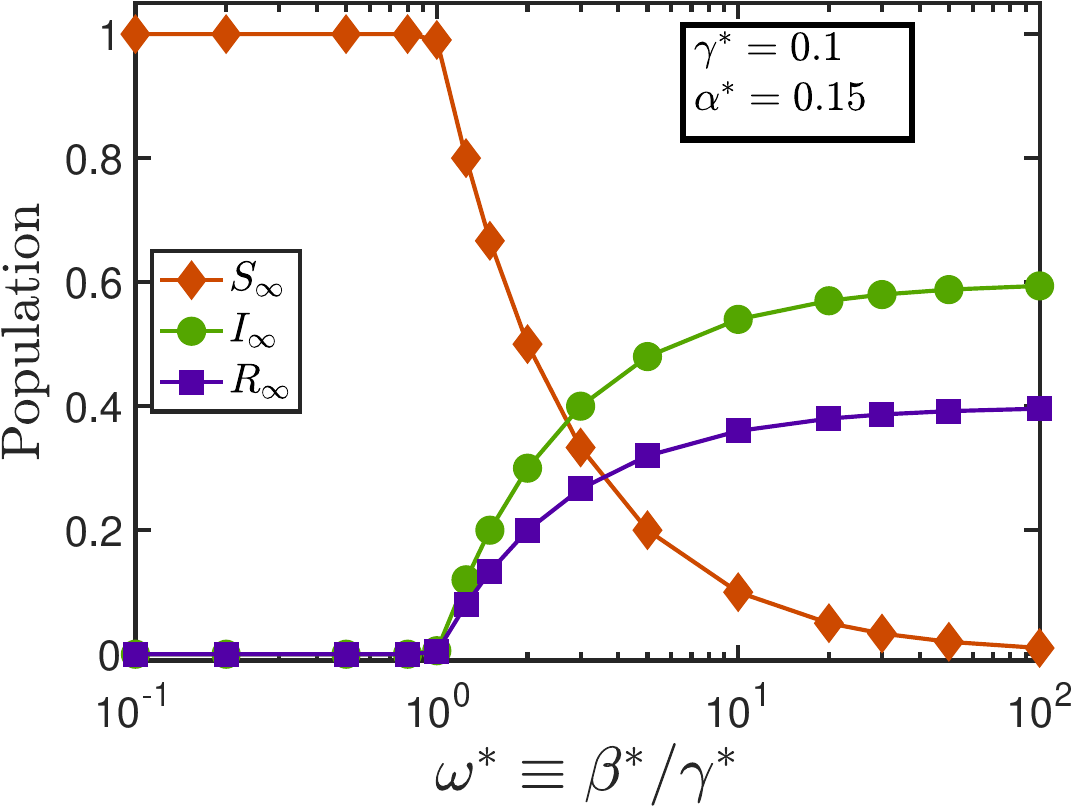}&
\includegraphics[width=2.1in,height=!]{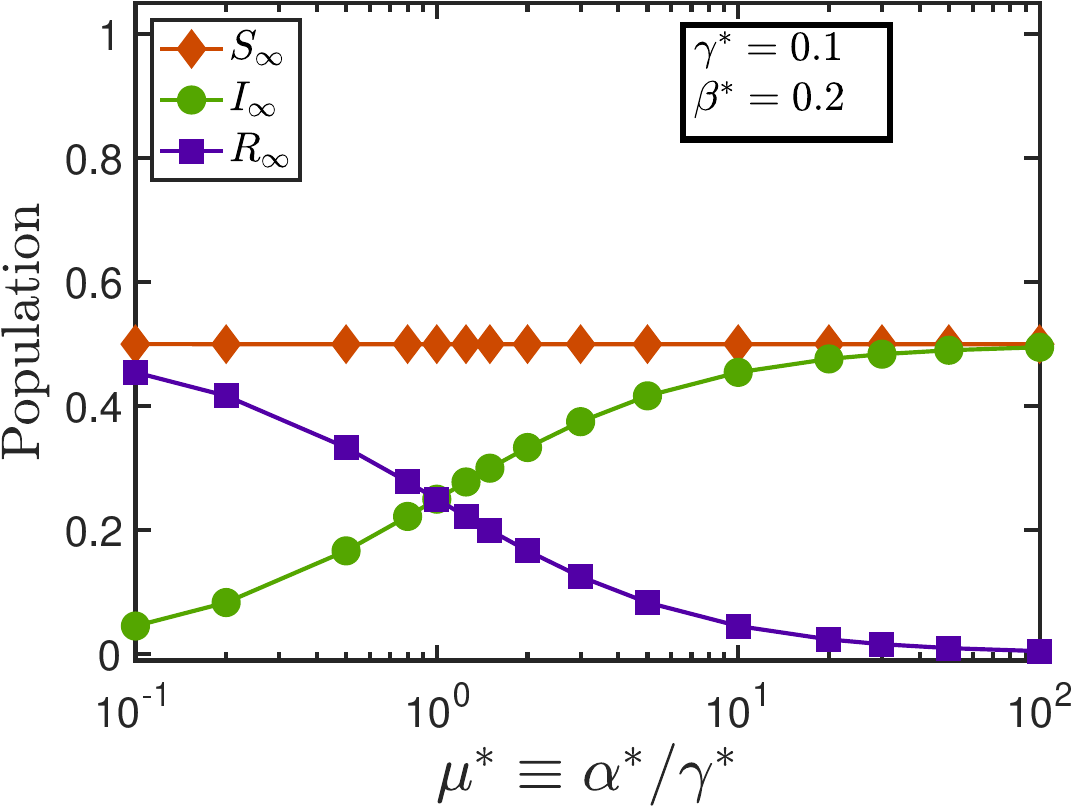}&
\includegraphics[width=2.1in,height=!]{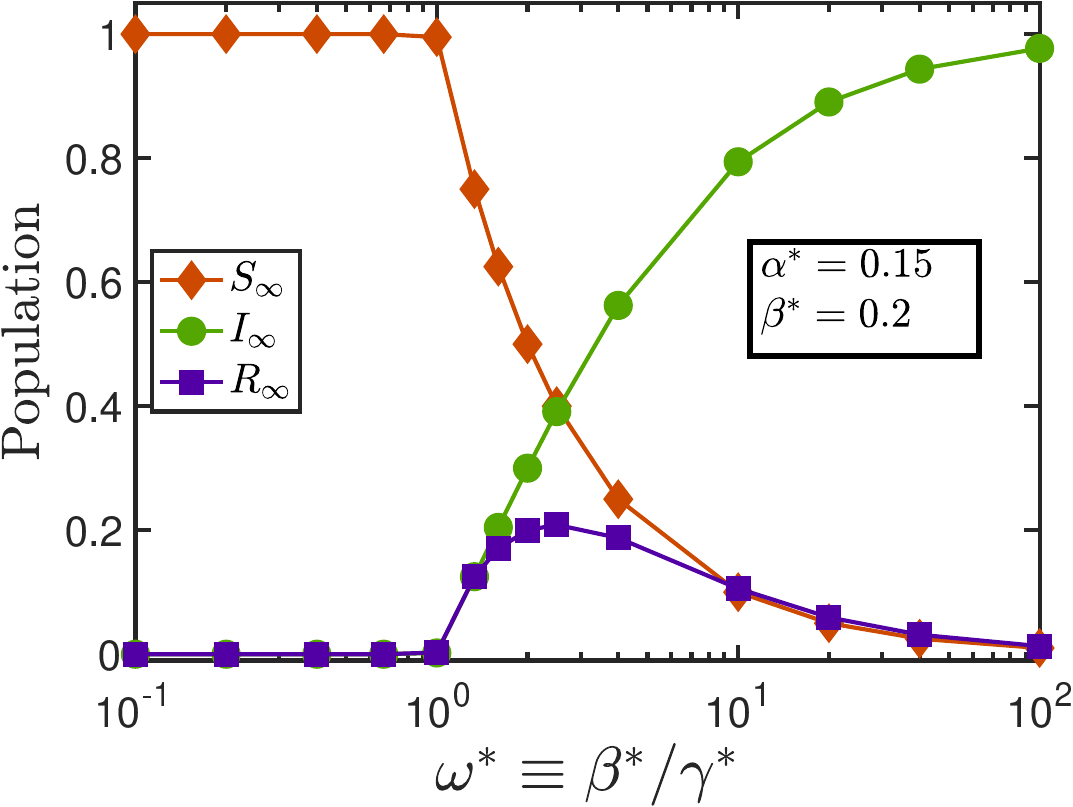}\\
(a) & (b) & (c)\\
\end{tabular}
\end{center}
\caption{Steady-state population as a function of (a) relative transmission rate at constant values of recovery and relapse rates, (b) relative relapse rate at constant values of infection and recovery rates, and (c) relative transmission rate at constant values of infection and relapse rates. Data in Figs.~\ref{fig:tevol_macro} and ~\ref{fig:stead_pop} have been generated by solving the coupled ODEs given by eq.~(\ref{eq:macro_eqn}), using a code shared on Mathworks File Exchange~\cite{Valentini2020}.}
\label{fig:stead_pop}
\end{figure}

~\citet{Norambuena2020} simulated a collection of active Brownian particles as self-propelled agents moving in two dimensions in which a susceptible particle gets infected as soon as it comes within a cut-off distance of an infected particle. They considered a low particle density system, meaning that the number of active agents per unit dimensionless area was approximately $0.03$, and used a mean-free path approach to determine analytical expressions for the rate of infection in their model. The internal state of the particle does not affect its dynamics, and all particles move with the same self-propulsion speed. ~\citet{Zhao2022} also endow an internal state to active agents and examine how the effectiveness of the disease spread differs from the macroscopic model that assumes a well-mixed population. Here too, a particle's self-propulsion speed is unaltered by its internal state. Useful future directions of research identified in their paper include the study of models in which the internal state is coupled to the motility of the particle, and the simulation of systems at a large enough area fraction so that MIPS can be observed. ~\citet{Forgacs2022} study contagion dynamics using an agent-based modeling approach, where the state of the particle determines its motility, i.e., the infected particles move $50\%$ slower than susceptible or recovered particles. Their simulation considers a collection of active Brownian disks moving in two dimensions, the majority of which belong to a large cluster that has undergone MIPS. The evolution of the disease for various values of the epidemiological constants is examined, along with a characterization of the spatial distribution of the susceptible particles around the infected particles. The dynamics of active particles that exchange information about their internal state has been studied in contexts outside of epidemiological modeling as well~\cite{Paoluzzi2020,Zhong2023}. For example, ~\citet{Paoluzzi2020} examine MIPS in a mixture of motile and non-motile particles that can change their identities upon collision. Quantifying phase separation arising in agent-based modeling of epidemiological systems remains an open question, particularly when the internal state is coupled to the motility of the agents, causing the system to behave as a transient mixture of active and passive particles. Also pertinent is establishing the connection (or lack thereof) between the macroscopic and microscopic descriptions of disease spread in populations, in the non-dilute limit. We address these questions in the present work and find that: (a) the analytical theory defining the phase boundary in a system of active and passive particles~\cite{Stenhammar2015} can successfully be adapted to describe phase separation in a collection where the activity of the particles switches \textit{transiently}, and (b) certain qualitative similarities can be observed in the disease statistics predicted by the macroscopic and microscopic models, although a direct mapping between the two is not found.

The remainder of this manuscript is organized as follows: Sec.~\ref{sec:mac_model} recapitulates salient features of the macroscopic SIRS model, Sec.~\ref{sec:sim_method} describes the numerical implementation of the microscopic model for two protocols of the spread of infection: one in which each infected particle can pass the disease on to exactly one susceptible particle within a contagion radius, and another in which an infected particle can potentially spread the disease to multiple susceptible particles within the contagion radius. A comparison of the contagion dynamics predicted by the one-to-one and one-to-many infection protocols are presented in Sec.~\ref{sec:ab_comparison}. Sec.~\ref{sec:macro_micro} presents a comparison of the disease dynamics obtained using the macroscopic and microscopic models. A discussion of phase separation in a mixture of active and transiently passive disks is presented in Sec.~\ref{sec:phase_diag}, accompanied by a description of the protocol used in the present work for identifying the occurrence of phase separation in such systems, which could be used to analyze MIPS in situations beyond the present study. We conclude in Sec.~\ref{sec:concl}.

\begin{figure}
\begin{center}
\includegraphics[width=7cm,height=!]{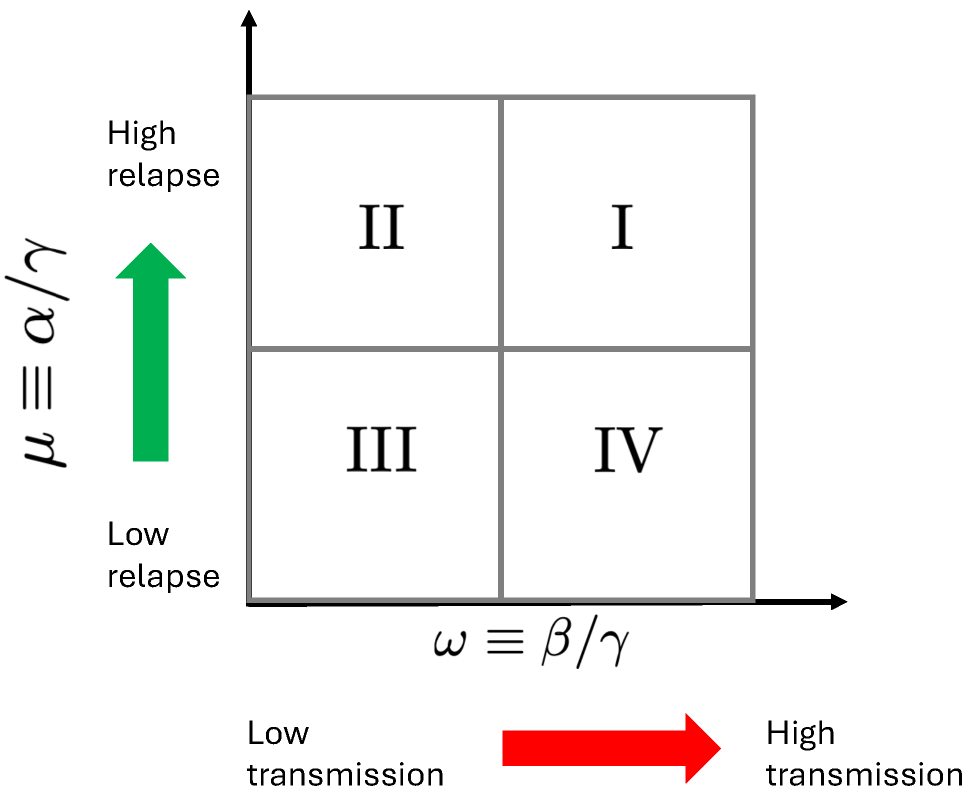}
\end{center}
\caption{Schematic of parameter space explored. {Parameters in quadrant I correspond to the case of high infectiousness (or transmission), and high loss of immunity (relapse), while quadrant III corresponds to the case of low infectiousness and low relapse. Other quadrants could interpreted in a similar manner.}}
\label{fig:par_space}
\end{figure}

\begin{figure}[t]
\begin{center}
\begin{tabular}{c c}
\includegraphics[width=7cm,height=!]{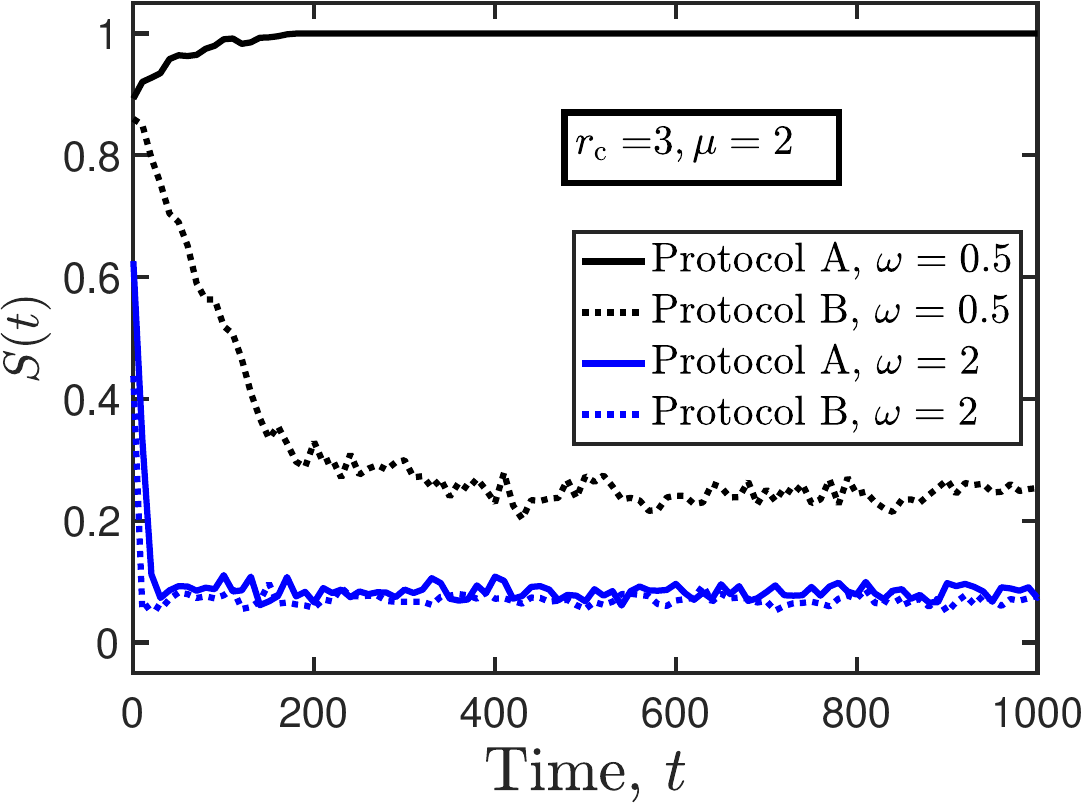}&
\includegraphics[width=7cm,height=!]{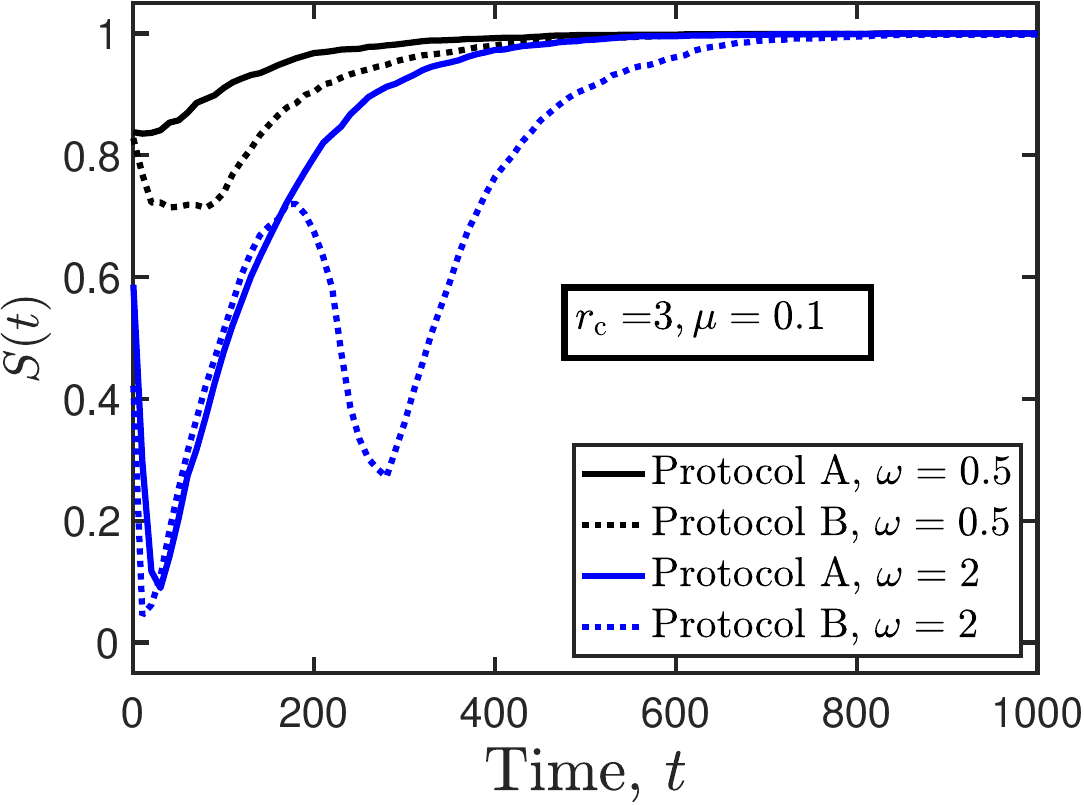}\\
(a) & (b)\\
\end{tabular}
\end{center}
\caption{Time-evolution of the susceptible population predicted by the two protocols at (a) high and (b) low values of the relative relapse rate $\mu\equiv\alpha/\gamma$, for relative transmission rates of $\omega=0.5$ and $\omega=2$.}
\label{fig:ab_transient}
\end{figure}

\section{\label{sec:mac_model} Macroscopic model revisited}

We consider the SIRS model in this work, in which the immunity gained by the recovered particles can be lost, causing them to become susceptible to the infection again. {The terms ``macroscopic" model and ``well-mixed" model are used interchangeably, to refer to the set of ODEs given by eq.~(\ref{eq:macro_eqn})}. This model is typically used to capture the contagion dynamics of diseases like influenza~\cite{Edlund2011,Bucyibaruta2023} and the Omicron variant of SARS-CoV-2~\cite{Nill2023}, where the immunity gained is temporary. The population of each compartment is denoted by $N_S$, $N_I$ and $N_R$, and evolve according to the following coupled ODEs:
\begin{equation}\label{eq:macro_eqn}
\begin{split}
\dfrac{dN_{S}}{dt}&=-\dfrac{\beta^{*}N_{I}N_{S}}{N}+\alpha^{*} N_{R},\\[5pt]
\dfrac{dN_{I}}{dt}&=\dfrac{\beta^{*}N_{I}N_{S}}{N}-\gamma^{*} N_{I},\\[5pt]
\dfrac{dN_{R}}{dt}&=\gamma^{*} N_{I}-\alpha^{*} N_{R}
\end{split}
\end{equation}
where $\beta^{*}$ indicates the rate of infection or transmission of the disease from an infected to a susceptible individual, $\gamma^{*}$ represents the rate of recovery of an infected individual, and $\alpha^{*}$ represents the rate of relapse or the loss of immunity of a recovered individual, resulting in its conversion to a susceptible agent. We also define the relative transmission rate $\omega^{*}\equiv\beta^{*}/\gamma^{*}$ and relative relapse rate $\mu^{*}\equiv\alpha^{*}/\gamma^{*}$ for convenience. The asterisks on the macroscopic (population-level) rate constants are used to distinguish them from their counterparts in the microscopic model, which is discussed later in the paper. 

\begin{figure}[t]
\begin{center}
\begin{tabular}{c c}
\includegraphics[width=7cm,height=!]{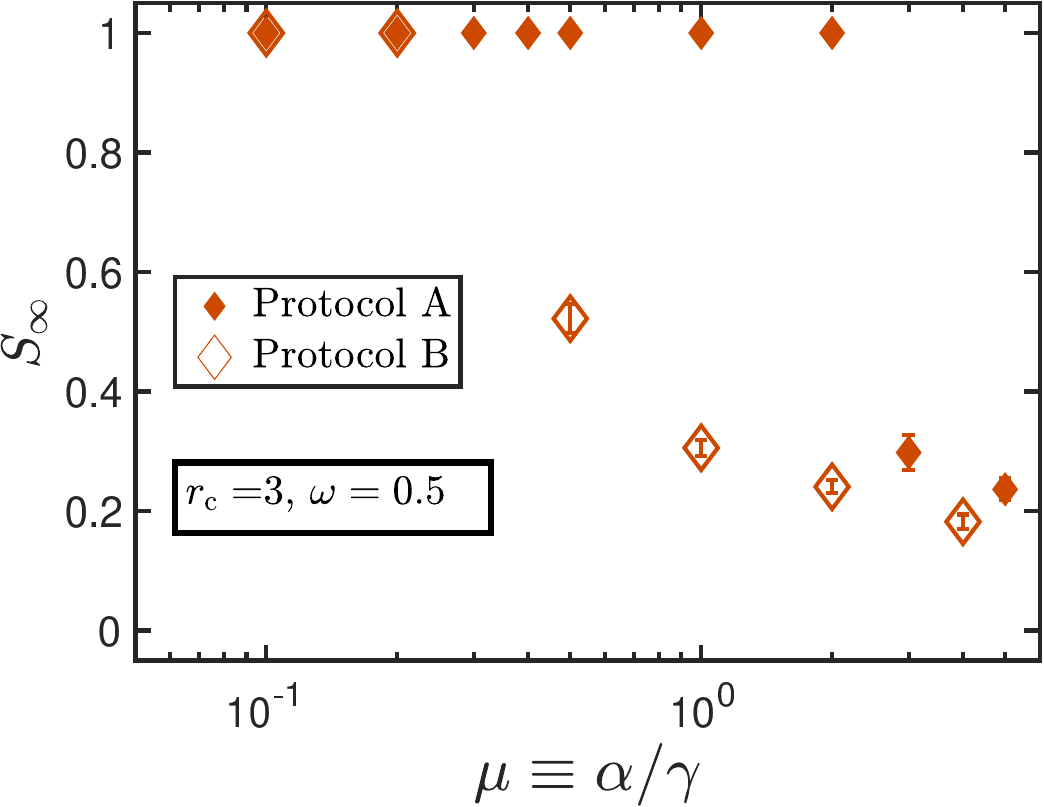}&
\includegraphics[width=7cm,height=!]{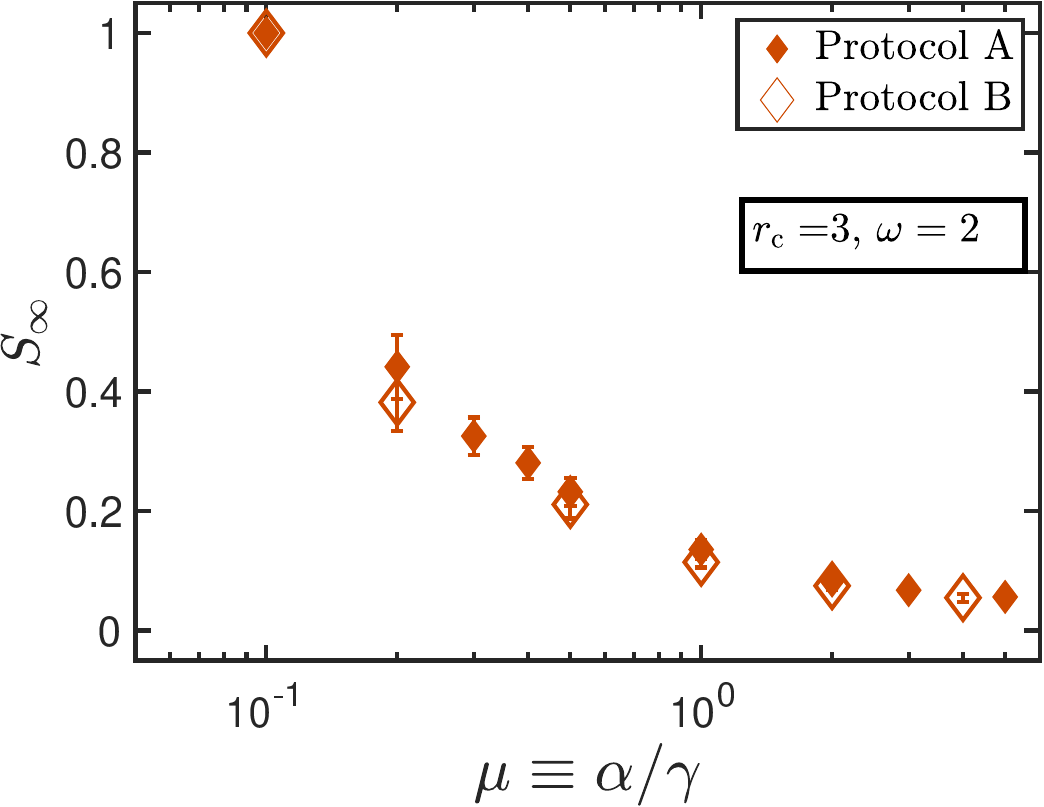}\\
(a) & (b)\\
\end{tabular}
\end{center}
\caption{Steady-state susceptible population as a function of the relative relapse rate $\mu$ for (a) high and (b) low values of the relative transmission rate $\omega$. Error bars represent standard deviation of data obtained from the time-averaging procedure used to estimate the mean steady-state population. Where invisible, error bars are smaller than symbol size.}
\label{fig:ab_steady}
\end{figure}

Depending on the context, the labels $\{S,I,R\}$ could refer to the type of population being discussed, or the normalized value of that population, e.g., $S=N_{S}/N$. The governing equations [Eq.~(\ref{eq:macro_eqn})] are subject to the initial conditions $S_{0}\equiv S(t=0),I_{0},R_{0}$, and the normalization condition 
\begin{equation}\label{eq:nmlz}
S+I+R=1.
\end{equation}
 Two possible steady-state solutions to the system of equations~(\ref{eq:macro_eqn}) emerge: the first corresponds to epidemic extinction, with $S_{\infty}=1,I_{\infty}=0,R_{\infty}=0$, and the second is
\begin{equation}\label{eq:steady_macro}
\begin{split}
S_{\infty}&=\dfrac{1}{\omega^{*}}\\[5pt]
I_{\infty}&=\left(1-S_{\infty}\right)\left[\dfrac{\alpha^{*}}{\alpha^{*}+\gamma^{*}}\right]\\[5pt]
R_{\infty}&=\left(1-S_{\infty}\right)\left[\dfrac{\gamma^{*}}{\alpha^{*}+\gamma^{*}}\right]
\end{split}
\end{equation}
Values of $\omega^{*}<1$ correspond to epidemic extinction, while $\omega^{*}>1$ results in the endemic state given by eq.~(\ref{eq:steady_macro}). Interestingly, the steady-state susceptible population depends only on the rates of infection and recovery, while the populations of the other-two species ($I$ and $R$) depend on all the three rate constants.

Fig.~\ref{fig:tevol_macro} illustrates the effect of the rate constants on the time-evolution of the infection, for different initial populations of the various populations. The rate constants and the initial conditions uniquely determine the dynamics of population evolution. The steady-state value of the various populations ($S_{\infty}$, for instance), however, are solely determined by the ratios $\omega^{*}$ and $\mu^{*}$, and are independent of the initial conditions. In the language of dynamical systems theory~\cite{Strogatz2015}, eq.~(\ref{eq:steady_macro}) is an attracting state for $\omega^{*}>1$.

Fig.~\ref{fig:stead_pop} illustrates a plot of the steady-state populations as a function of the rate constants. Fig.~\ref{fig:stead_pop}~(a) illustrates the case of varying the infection rate fixed values of the recovery and relapse rate. Low values of the relative transmission rate, i.e.,  $\omega^{*}<1$ result in epidemic extinction, in which there are no infected particles in the long-time limit. As $\omega^{*}\geq 1$, however, the steady-state population of susceptibles declines with an increase in the infection rate. The transition of the steady-state numbers from the low $\omega^{*}$ branch to the high $\omega^{*}$ branch appears to follow a transcritical bifurcation~\cite{Strogatz2015}. Holding the infection and recovery rates constant, while varying the relapse rate, as shown in Fig.~\ref{fig:stead_pop}~(b), has no effect on the population of susceptibles. There are more recovered than infected individuals for values of $\mu^{*}<1$, while the balance is reversed as the value of $\mu^{*}$ crosses unity. Lastly, Fig.~\ref{fig:stead_pop}~(c) examines the consequences of varying the recovery rate as the infection and relapse rates are held constant. At both small and large values of the recovery rate, the recovered population is nearly zero. This is because for $\omega^{*}<1$ the population is solely made of susceptible individuals, while the infected individuals dominate the population for $\omega^{*}\gg1$. 

\section{\label{sec:sim_method} Numerical simulations of microscopic model for SIRS dynamics in active Brownian particles}

\begin{figure*}[t]
\begin{center}
\begin{tabular}{c c}
\includegraphics[width=2.5in,height=!]{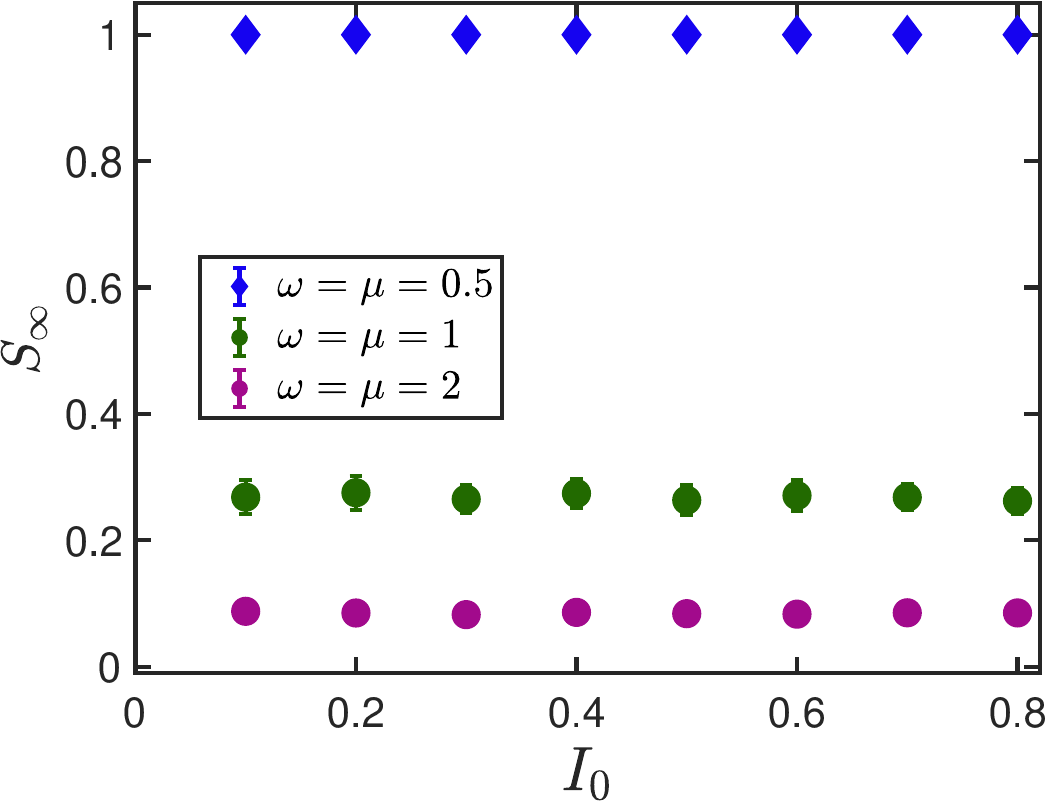}&
\includegraphics[width=2.5in,height=!]{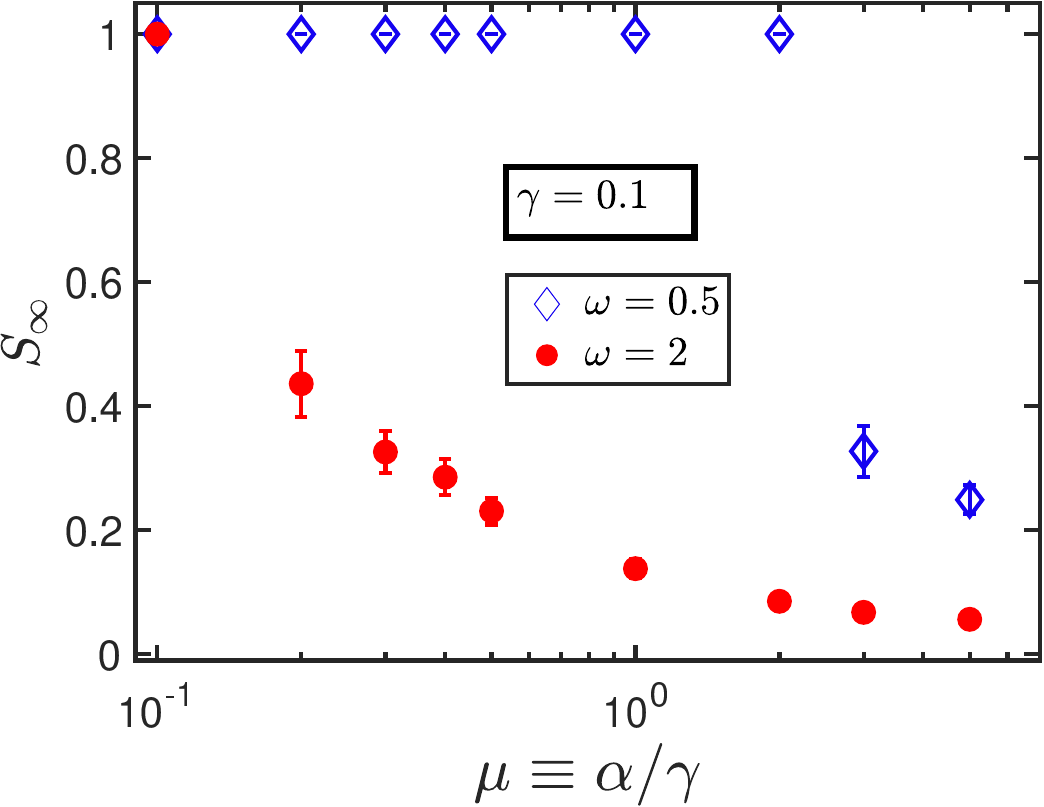}\\
(a) & (b) \\
\includegraphics[width=2.5in,height=!]{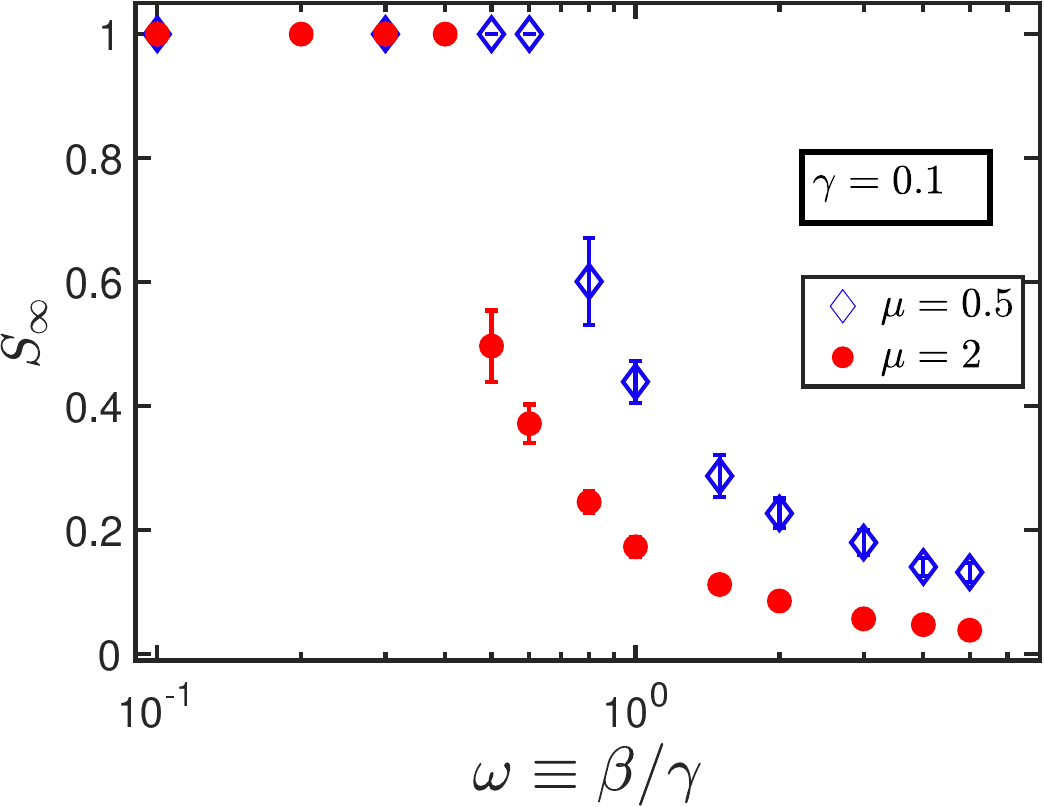}&
\includegraphics[width=2.5in,height=!]{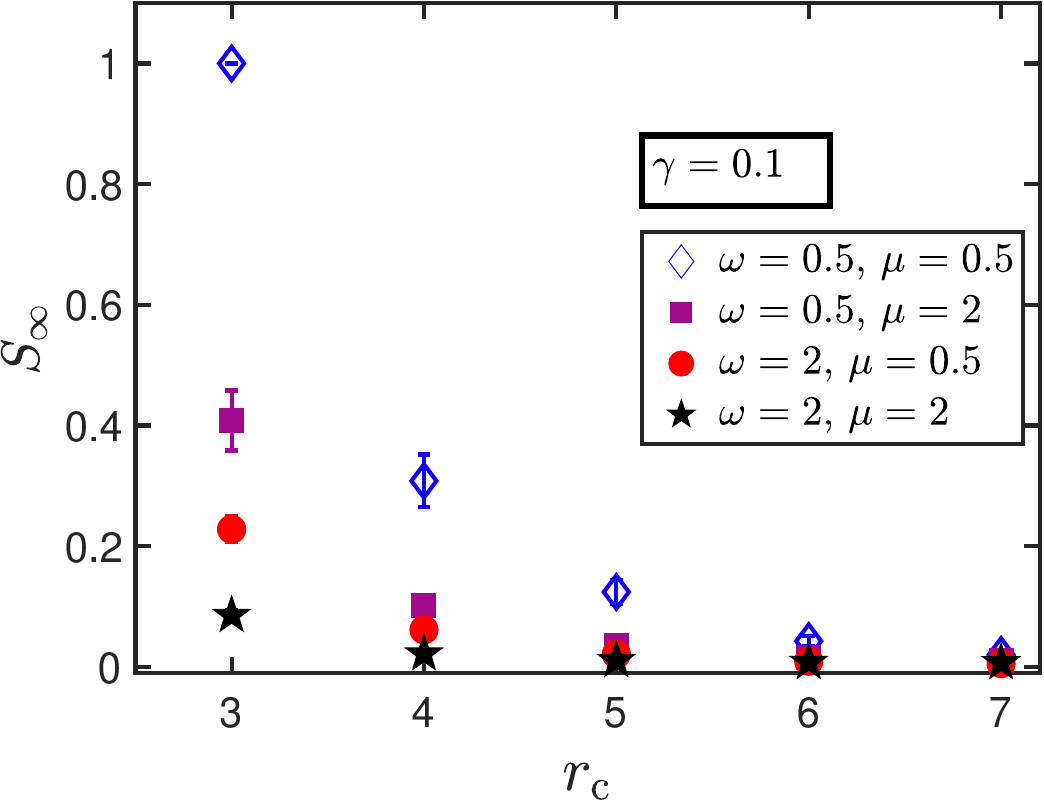}\\
(c) & (d)\\
\end{tabular}
\end{center}
\caption{Steady-state susceptible population as a function of (a) initial fraction of infected individuals, (b) relative relapse rate at fixed values of infection rate, (c) relative transmissibility at fixed values of relapse rate, and (d) contagion radius at fixed values of the epidemiological constants. All results obtained using Protocol A (one-to-one) for disease spread.}
\label{fig:micro_model}
\end{figure*}

We simulate a system of $N$ active Brownian particles (ABPs) of unit radius each ($d=2r=2$) moving in a periodic square box of side $L$. The position of the particles evolves in time according to
\begin{equation}\label{eq:pos_dyn}
\dot{\bm{r}}_{i}=\bm{v}^{\text{HM}}_{i}+U_{0}\bm{e}_{i}
\end{equation}
where $U_{0}$ is the self-propulsion speed of the disks. The particles move in the direction $\bm{e}_{i}\equiv\left[\cos\theta_{i},\sin\theta_{i}\right]$, where $\theta_{i}$ denotes the particle's orientation measured with respect to the positive $x-$axis. The particle orientation evolves in time according to a rotational diffusive process, such that
\begin{equation}\label{eq:r_noise}
\left<\dot{\theta}_{i}(t)\dot{\theta}_{i}(t')\right>=2D_{\text{r}}\delta(t-t')
\end{equation}
where $D_{\text{r}}$ denotes the rotational diffusion constant. The positions and orientations of the particles are updated using a forward Euler algorithm with a timestep $\Delta t$. {The $\bm{v}^{\text{HM}}_{i}$ term on the RHS of eq.~(\ref{eq:pos_dyn}) represents a harmonic interaction that operates only when the centre-to-centre separation of disk $i$ and $k$ is smaller than their diameter $d$. The steric interaction is proportional to the extent of overlap of the disks, and acts to alter the position of each disk in an overlapping pair such that they are just in contact. The functional form of $\bm{v}^{\text{HM}}_{i}$ is given by
\begin{equation}\label{eq:hm_vel}
\bm{v}^{\text{HM}}_{i}=\dfrac{1}{\Delta t}\sum_{k}^{n_{o}}K\left(r_{ik}-d\right)\Theta\left(d-r_{ik}\right)\hat{\bm{r}}_{ik},
\end{equation}
where $\hat{\bm{r}}_{ik}=\bm{r}_{ik}/r_{ik}$ is the unit vector along the line joining the particle centres, $\Theta$ denotes the Heaviside function, the stiffness $K=0.5$, and $n_{o}$ denotes the number of overlapping particles in the neighborhood of the $i^{\text{th}}$ particle.} 

We next describe the update rules to simulate the spreading of diseases in this microscopic, agent-based model, which largely follows the algorithm outlined by~\citet{Forgacs2022}. {The simulations are initialized in a homogeneous configuration, with all the agents arranged in regularly spaced intervals on a square lattice.} Each particle has an associated internal state, i.e., it could be a susceptible ($S$), infected ($I$), or recovered ($R$) agent. {The contagion dynamics (change in the internal state of the agent, i.e., S,I,R) as well as the physical movement of the agents happen simultaneously.} {This approach differs from that adopted by~\citet{Forgacs2022}, in which the contagion dynamics starts from a phase separated state.}

A census of the number of particles in each sub-category is taken at the beginning of each timestep. The allowed transitions are $S\to I$, $I\to R$ and $R\to S$, with the rate constants associated with the transitions given by $\beta,\gamma$ and $\alpha$, respectively. The protocol governing the spread of the infection is as follows: a loop is run over all the $N_I$ infected particles in the box at a given time instant, and the number of susceptible and infected particles within a cut-off radius $r_{\text{c}}$ of an infected particle $I^{(n)}$ is recorded as $N_{S}^{(n)}$ and $N_{I}^{(n)}$, respectively. Note that the index $n$ runs over all the infected disks in the system at a given time instant. By this definition, the smallest allowable value for $N_{S}^{(n)}$ is zero, while that for $N_{I}^{(n)}$ is unity. The probability of infection is calculated as $p_{\text{inf}}=\beta N_{I}^{(n)}\Delta t$. The probability of recovery is given by $p_{\text{rec}}=\gamma\Delta t$.  The values of the three probabilities $\{p_{\text{inf}},p_{\text{rec}},p_{\text{rel}}\}$ are capped at unity. If any of these probabilities exceed unity during a timestep, it is reset to unity.

\begin{figure}
\begin{center}
\includegraphics[width=8cm,height=!]{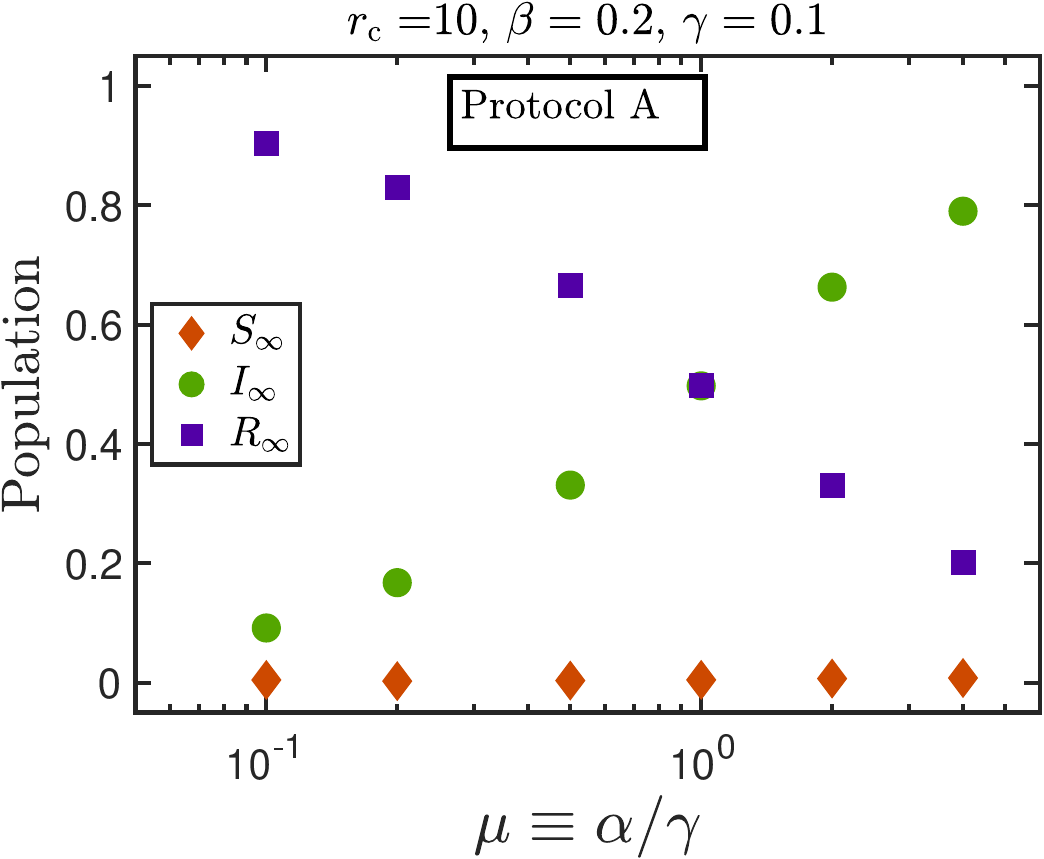}
\end{center}
\caption{Steady state population as a function of the relative relapse rate, as obtained using Protocol A.}
\label{fig:rc_10_nums}
\end{figure}

We consider two protocols by which the infection might spread: 

\textbf{Protocol A: one-to-one}

For each value of the index $n$ during the loop over $N_I$, if the outcome of a binomial trial with probability $p_{\text{inf}}$ is non-zero, then $one$ susceptible particle from $N_{S}^{(n)}$ is picked at random for conversion to an infected particle. Similarly, if the outcome of a binomial trial with probability $p_{\text{rec}}$ is non-zero, and there are multiple infected particles within a cut-off radius $r_{\text{c}}$ of an infected particle $I^{(n)}$, then the state of $one$ infected particle changes to recovered. An infected particle in this protocol can spread the disease to only one other particle in a timestep. Having multiple infected particles in the vicinity of a susceptible particle only increases the probability of infection.

\textbf{Protocol B: one-to-many}

For each value of the index $n$ during the loop over $N_I$, the number of susceptible particles that get infected is decided by drawing a sample from a binomial distribution of probability $p_{\text{inf}}$, $N_{S}^{(n)}$ number of times, and calculating $S^{+}$, the total number of successful outcomes. A total of $S^{+}$ particles from amongst $N_{S}^{(n)}$ are then randomly selected to be infected. Once the loop over all $N_I$ is completed, the number of infected particles that recover is decided by drawing a sample from a binomial distribution of probability $p_{\text{rec}}$, $N_{I}$ number of times, and calculating $I^{+}$, the total number of successful outcomes. A total of $I^{+}$ particles from amongst $N_{I}$ are then randomly selected to undergo recovery. In this protocol, therefore, an infected particle could potentially transmit the disease to multiple susceptible particles in its neighborhood.  

\begin{figure}[t]
\begin{center}
\begin{tabular}{c c}
\includegraphics[width=6.3cm,height=!]{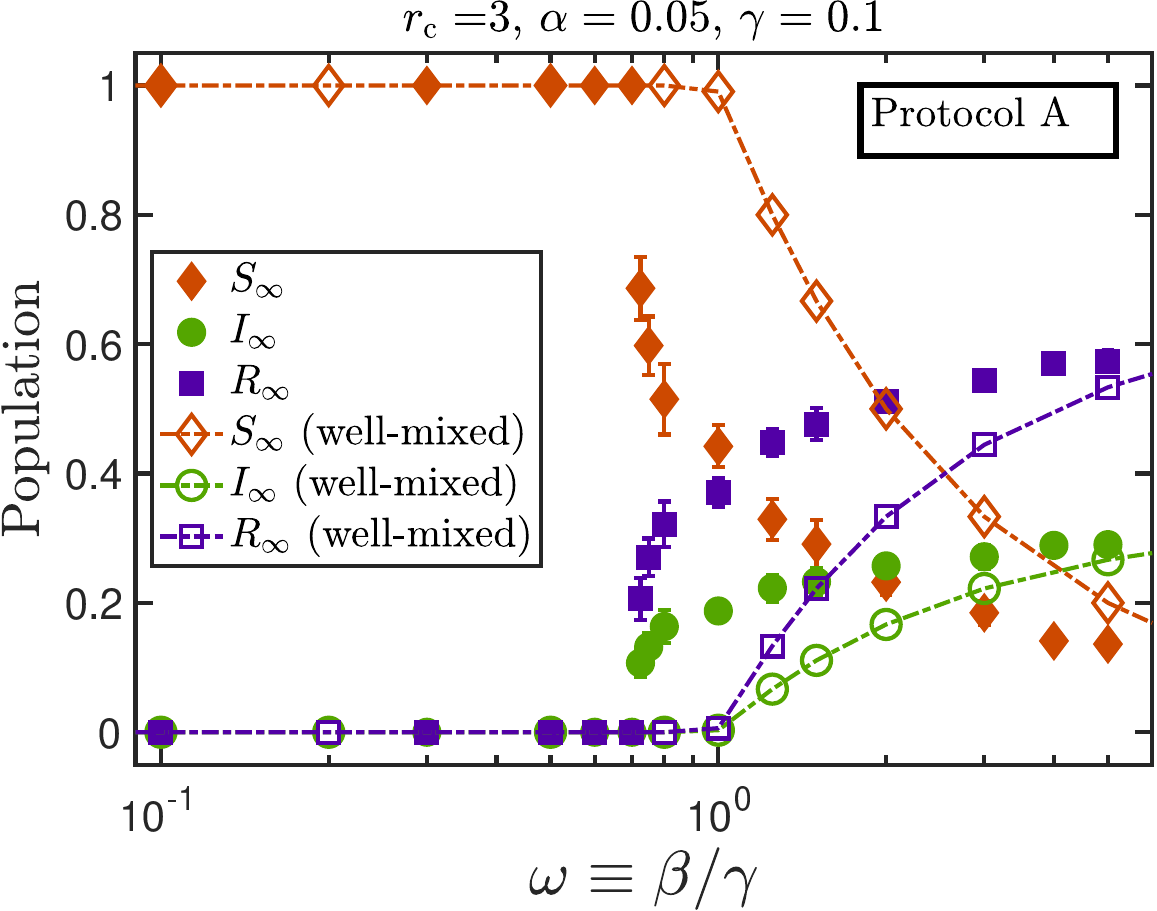}&
\includegraphics[width=6.3cm,height=!]{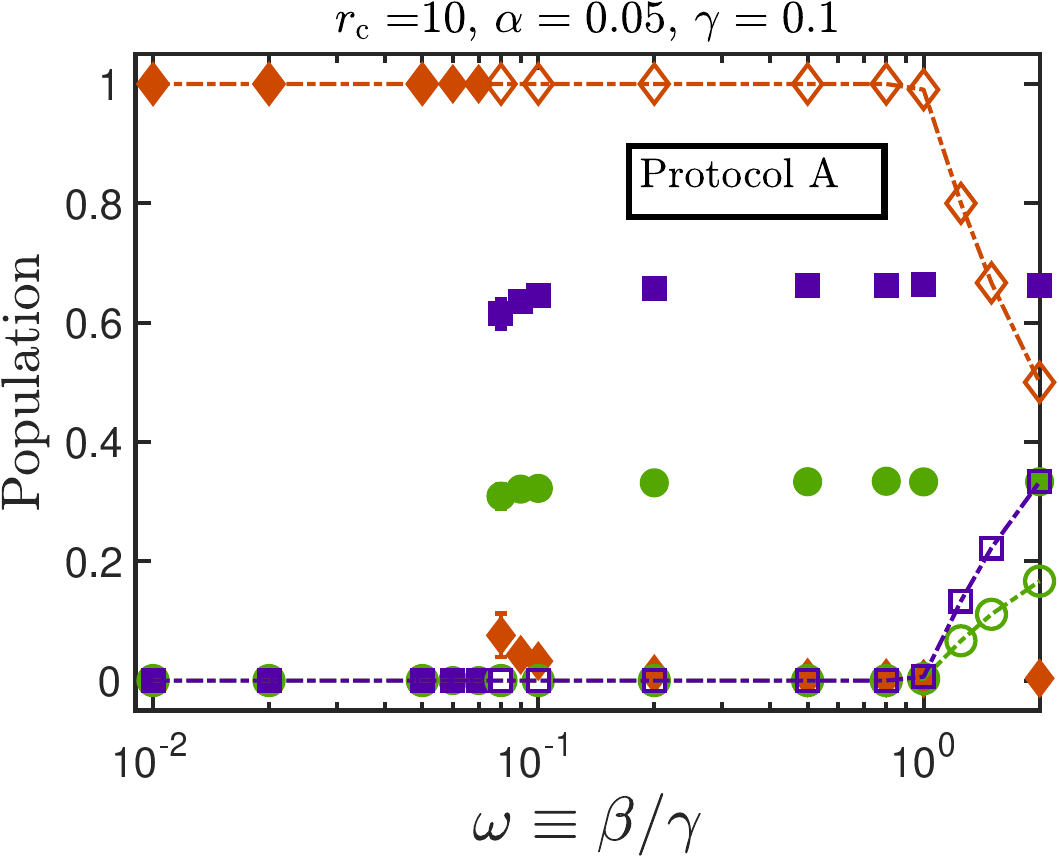}\\
(a) & (b)\\
\end{tabular}
\end{center}
\caption{{Steady-state epidemic statistics as a function of the relative infection rate, at a contagion radius of (a) $r_{\text{c}}=3$ and (b) $r_{\text{c}}=10$. The same legend scheme is followed in both the subfigures, with filled symbols denoting the microscopic model results, obtained using Protocol A, and hollow symbols representing the well-mixed model results. The numerical values of the parameters used in both the models are the same, although there is no direct mapping between the two.}}
\label{fig:micro_model_alph}
\end{figure}

For both the protocols discussed above, the relapse of recovered particles into the susceptible category is governed by a binomial process of probability $p_{\text{rel}}=\alpha\Delta t$, and is invoked once the loop over all $N_I$ is completed. Furthermore, the infected particles become immobile ($U_{0}=0$), while both the susceptible and recovered particles retain their activity. The infected particles regain their mobility upon recovery. The instantaneous fraction of active disks in the system is therefore given by $x_{A}\equiv\left(N_{S}+N_{R}\right)/N$. 

The population and system size in all cases are chosen to be $N=1600$ and $L=100$, corresponding to a number density of $\rho=N/L^2=0.16$, and an area fraction of $\phi_{0}=\rho\pi d^2/4\approx 0.5$. The self-propulsion speed is fixed at $U_{0}=0.1$. A discrete timestep width of $\Delta t=0.1$ is used in all the simulations, such that the displacement of a disk over a single timestep is smaller than its diameter. The contagion radius is chosen to be $r_{\text{c}}=3$ in all the simulations, unless specified otherwise. The motility of the system is quantified using the P\'{e}clet number, $\text{Pe}\equiv 3U_{0}/dD_{\text{r}}$. The persistence length $\ell\equiv U_{0}/D_{\text{r}}$ of the active particles must be smaller than the box dimensions, to minimize finite-size effects~\cite{Patch2017,Patch2018}. We adjust the rotational diffusion constant at each $\text{Pe}$ so that this condition is met. The effects of thermal noise on the translational motion are ignored in the present work, but may be included by adding a white-noise process to the RHS of eq.~(\ref{eq:pos_dyn}). In Section~\ref{sec:ab_comparison} that compares the contagion dynamics predicted by the two protocols (A and B), and in Section~\ref{sec:macro_micro} that explores the connection between the microscopic and macroscopic models, the numerical results are obtained from simulations that are $n=2\times 10^{4}$ steps long. In Section~\ref{sec:phase_diag} that charts the phase behavior of a collection of active and transiently passive disks, simulations of at least $\mathcal{O}(10^5)$ steps are used. The total length of the simulation is simply the product of the number of simulation steps and the discrete timestep, and is denoted by $t_{\text{sim}}=n\Delta t$.

\section{\label{sec:ab_comparison} Comparison between protocols A and B}

In this section, we compare the dynamics of infection spread, and the steady-state statistics for numerical simulations performed using the two protocols (A and B) discussed above. Given the broad span of the parameter space, we restrict ourselves to studying two values of the relative transmission rate $\omega\equiv\beta/\gamma$, and perform a scan across a range of relative relapse rates, $\mu\equiv\alpha/\gamma$, by holding the recovery rate constant at $\gamma=0.1$, and varying $\alpha$ and $\beta$ appropriately. We divide the parameter space into four quadrants as shown in Fig.~\ref{fig:par_space} and examine the simulation results accordingly.

\begin{figure}[t]
\begin{center}
\includegraphics[width=8cm,height=!]{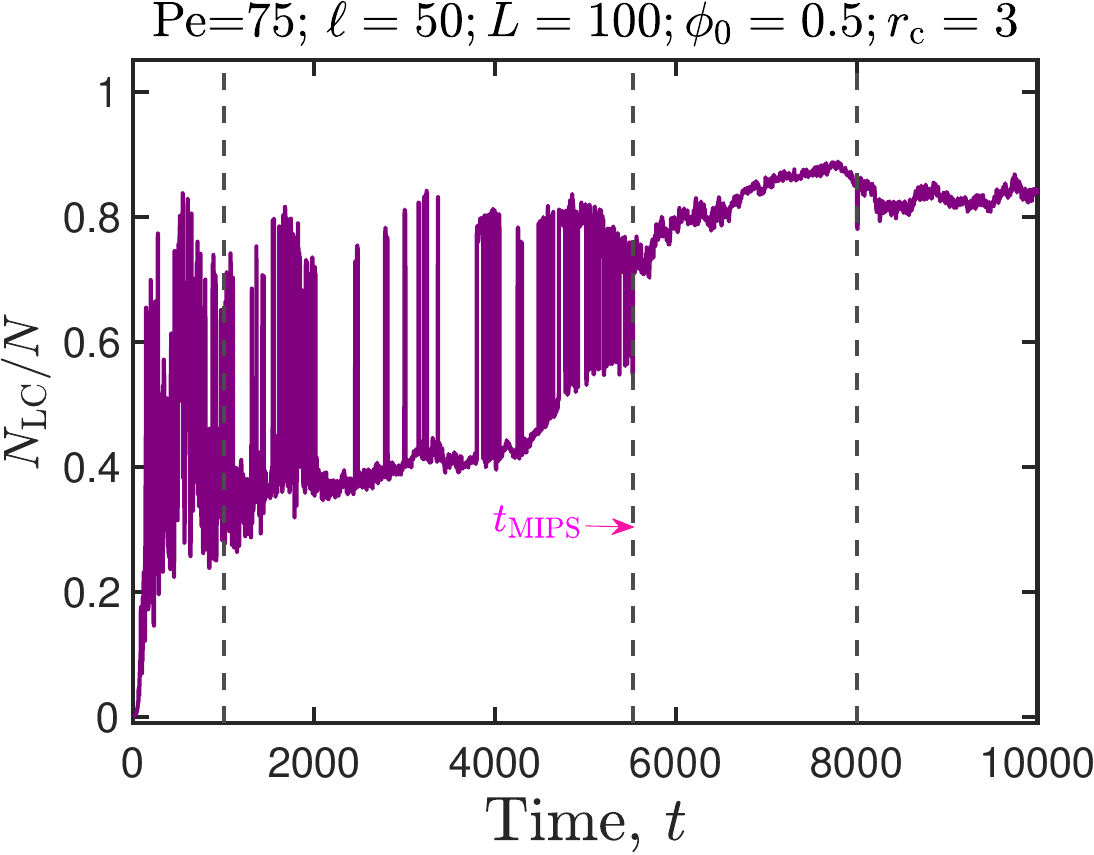}
\end{center}
\caption{Normalized size of the largest cluster as a function of time, for the case of $N=1600$ active disks in which all the epidemiological constants have been set to zero and there is no contagion dynamics. Snapshots of the system at the three indicated time-instances are given in Fig.~\ref{fig:snapshots_base_case_mips}.}
\label{fig:base_case_mips}
\end{figure}

\begin{figure*}[h]
\begin{center}
\begin{tabular}{c c c}
\includegraphics[width=4cm,height=!]{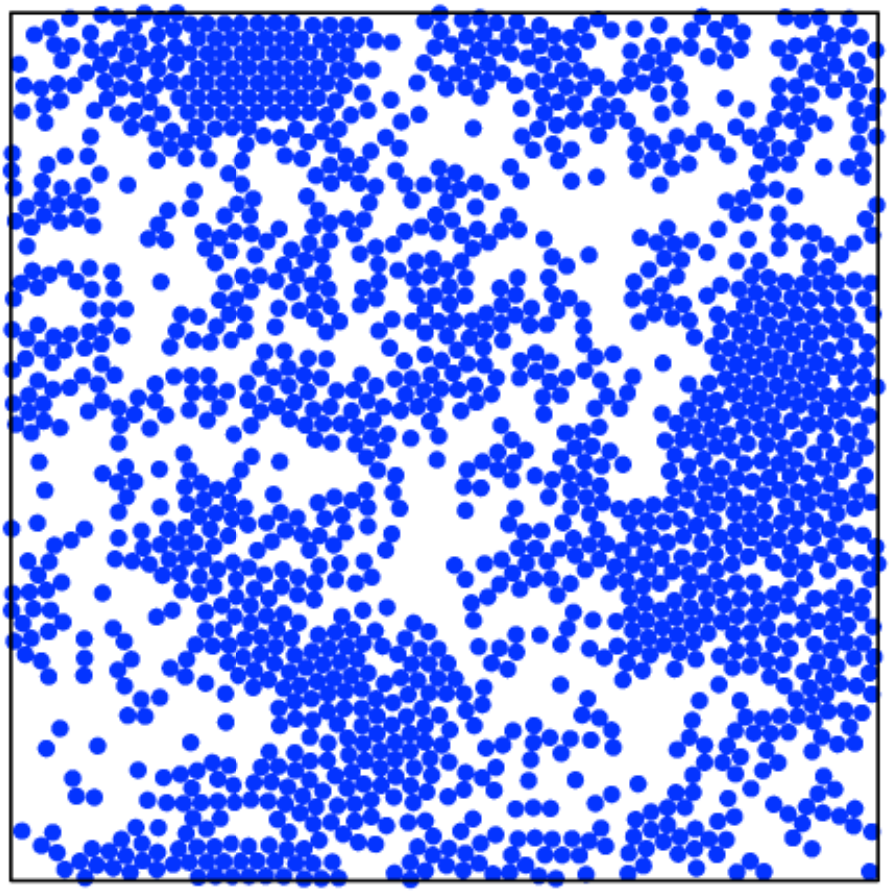}&
\includegraphics[width=4cm,height=!]{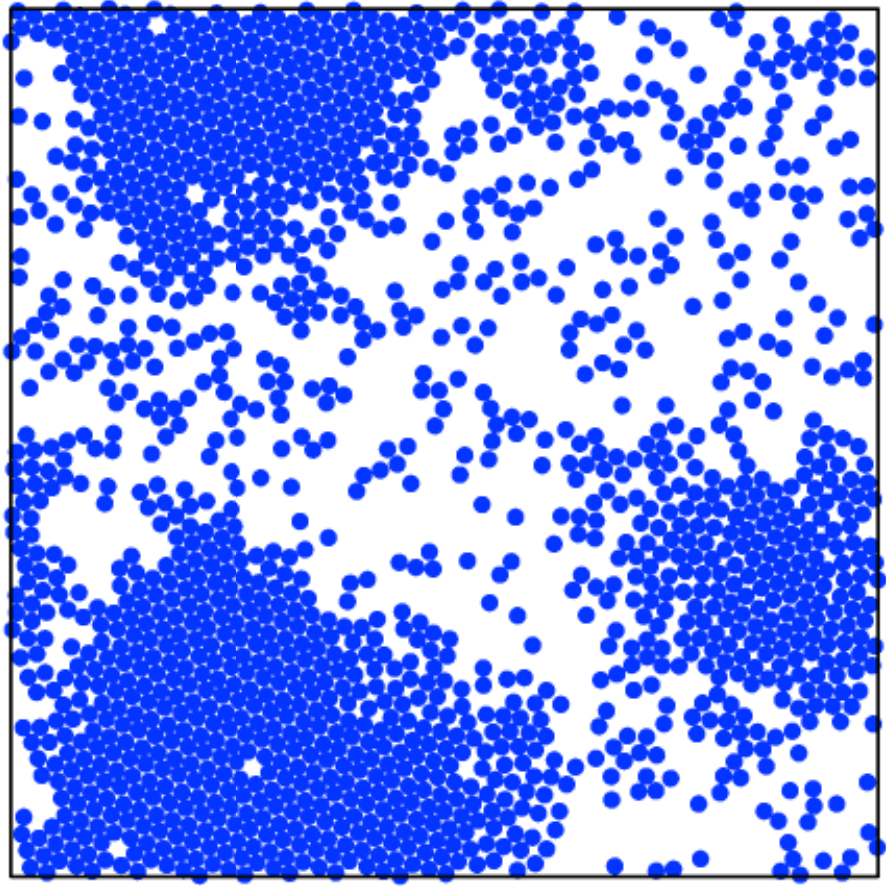}&
\includegraphics[width=4cm,height=!]{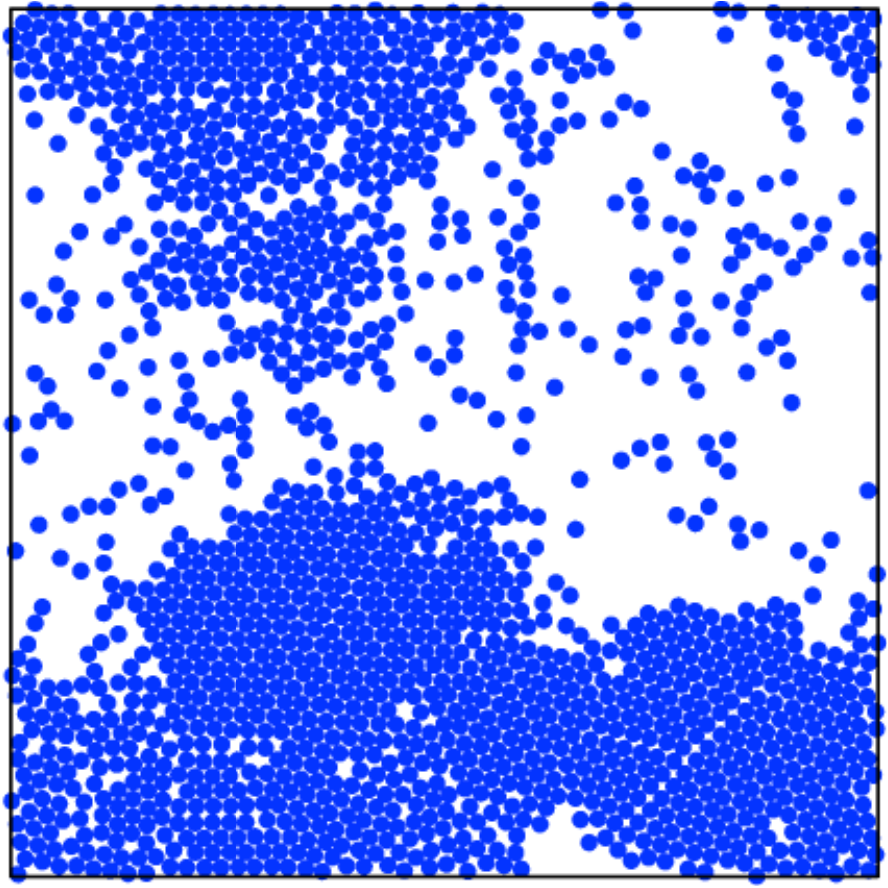}\\
(a)&(b)&(c)\\
\end{tabular}
\end{center}
\caption{Snapshots of the system in Fig.~\ref{fig:base_case_mips}, recorded at (a) $t=1000$, (b) $t=5527$, and (c) $t=8000$, respectively.}
\label{fig:snapshots_base_case_mips}
\end{figure*}

Fig.~\ref{fig:ab_transient}~(a) explores the first and second quadrants of the parameter space and illustrates that, at a higher value of the relative relapse rate, $\mu=2$, we notice two different behaviors, based on the value of the relative transmission rate. In Quadrant I, both the relative transmission and the relapse rates are high, implying that the disease spread is more probable, as is the replenishment of the numbers of the susceptible population. In this quadrant, we observe that the higher transmission rate plays a more dominant role, leading to a faster spread of the disease, and a rapid decrease in the number of susceptibles. Furthermore, it is immaterial if the infection spreads via the one-to-one or the one-to-many route, as they both result in a nearly identical prediction for the steady-state susceptible population. In Quadrant II, the relapse rate is higher than the infection rate. The steady-state in such a case is decided by the protocol for disease spread. Stipulating a one-to-one spreading protocol causes the infection to die out faster, so that the population is entirely composed of susceptibles in the long time limit. Allowing for a one-to-many spreading protocol for the infection compensates for the low value of the transmissibility in this quadrant, resulting in a faster spread of the disease, and a lower value of the steady-state susceptible population as compared to the one-to-one protocol. Fig.~\ref{fig:ab_transient}~(b) explores the third and fourth quadrants of the parameter space, for the smallest value of the relative relapse rate considered in this section, $\mu=0.1$. Over the time window considered in the figure, the numbers predicted by the one-to-many protocol are comparable to or lower than that predicted by the one-to-one protocol, indicating, unsurprisingly, that the former is more efficient in spreading the disease. This effectiveness of the one-to-many protocol, however, is manifest only transiently, as the steady-state values of the susceptible population are independent of the relative transmissibility, and the route for disease spread. 

From the above analysis, it is clear that the sharpest contrast between the steady-state outcomes predicted by the two protocols occurs in Quadrant II. Given the fluctuations in the values of the susceptible population, the average steady-state value, $S_{\infty}$, is estimated by computing the mean of the last $20\%$ of the time series. Fig.~\ref{fig:ab_steady} illustrates the steady-state susceptible population as a function of the relative relapse rate, for two different values of the relative transmission rate. The effectiveness of the one-to-many protocol in governing the spread of infection is most evident from the low transmission regime as identified in Fig.~\ref{fig:ab_steady}~(a). For $\mu>0.2$, the $S_{\infty}$ resulting from the one-to-many protocol is significantly lower than that predicted by the one-to-one protocol, indicating a spread of the disease amongst a larger fraction of the population. In the high transmission regime, however, the protocol for disease propagation has a less pronounced effect on the steady-state statistics, as evinced by Fig.~\ref{fig:ab_steady}~(b). Having compared the outcomes from the two protocols, we now present results obtained with Protocol A only for the rest of the paper.

\section{\label{sec:macro_micro} Connection between microscopic and macroscopic models}

A major distinction between the microscopic agent-based model (ABM) and the macroscopic population-based model for disease spread is the specification of a contagion radius $r_{\text{c}}$ for the former, which makes the spread of the infection depend not only on the number of infected and susceptible individuals at a given time, but also on their locations. We illustrate below a few salient features of the steady-state numbers predicted by the microscopic model, and how they compare to the macroscopic model predictions.

Fig.~\ref{fig:micro_model}~(a) illustrates that the steady-state population for the microscopic model is independent of the initial fraction of the infected population $I_{0}$, at a fixed value of the contagion radius ($r_{\text{c}}=3$), for various values of the relative transmission and relapse rates. The independence from initial conditions, over the range examined in this figure, is a trait shared by the microscopic and macroscopic models. 

Fig.~\ref{fig:micro_model}~(b) shows that the steady-state susceptible population decreases as a function of the ratio of the relative relapse rate ($\mu$), for fixed values of $\omega$ and the contagion radius ($r_{\text{c}}=3$). This marks a crucial departure from the macroscopic model (see fig.~\ref{fig:stead_pop}~(b)) in which $S_{\infty}$ is solely a function of the relative transmission rate.

The variation of the steady-state population as a function of the relative transmission rate $\omega$ is shown in Fig.~\ref{fig:micro_model}~(c), for fixed values of $\mu$ and contagion radius ($r_{\text{c}}=3$). The susceptible population is independent of the relative transmission rate for small values of the latter. Beyond a threshold value of the relative transmission rate, however, the susceptible population decreases with $\omega$. The crossover value depends on the ratio $\alpha/\gamma$, unlike in the macroscopic model where the transition occurs at $\omega^{*}=1$ and is independent of the relapse rate. While the existence of a bifurcation in the macroscopic model predictions (a system of coupled ODEs) is unsurprising~\cite{Strogatz2015}, it is remarkable that an evidence of bifurcation is also seen in the agent-based model. This also indicates that the stochastic update rule for the various compartments is faithful to the contagion dynamics as predicted by the ordinary differential equations of the macroscopic model.

Lastly, Fig.~\ref{fig:micro_model}~(d) illustrates the dependence of $S_{\infty}$ on the contagion radius, for various values of the epidemiological rate constants. For all the cases examined in the figure, the steady-state susceptible population decreases as a function of $r_{\text{c}}$.

\begin{figure}[t]
\begin{center}
\includegraphics[width=8.3cm,height=!]{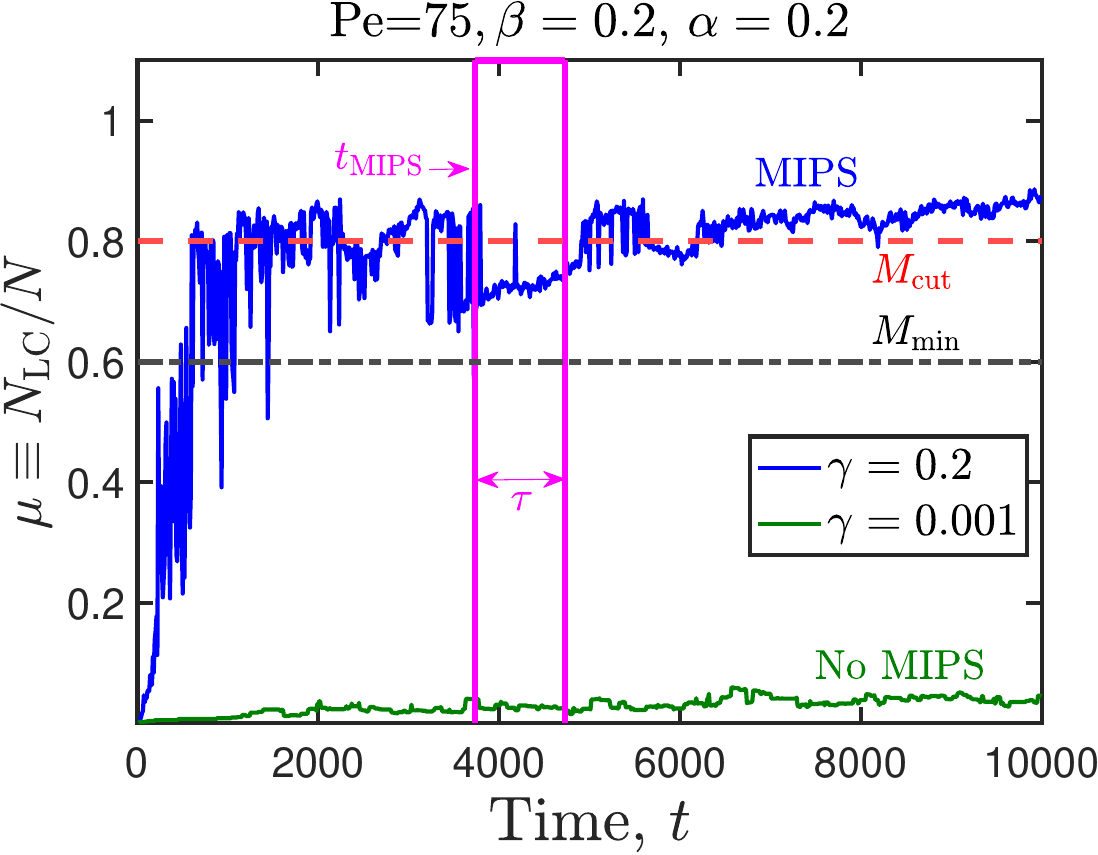}
\end{center}
\caption{Cluster size evolution as a function of time for two representative systems. Parameters needed for the algorithm that detects motility induced phase separation are identified here, and their numerical values provided in Table~\ref{param_MIPS}.}
\label{fig:mips_algo_1}
\end{figure}

\begin{table}[b]
\setlength{\tabcolsep}{12pt}
\centering
\caption{\label{param_MIPS} Parameters used in the algorithm (Fig.~\ref{fig:mips_algo_2}) to determine if phase separation has occurred or not.}
\begin{tabular}{|c| c|}
\hline
Parameter & Value used in present work\\
\hline
$M_{\text{cut}}$ & $0.8$ \\\hline
$M_{\text{min}}$ & $0.6$ \\\hline
$\tau$ & $0.1\,t_{\text{sim}}$\\\hline
$\sigma$ & $0.1$ \\ \hline
\end{tabular}
\end{table}

We briefly revisit the comparison between protocols A and B before concluding this section. The results reported in Section~\ref{sec:ab_comparison} all come from numerical simulations with a contagion radius of $r_{\text{c}}=3$. Fig.~\ref{fig:rc_10_nums} illustrates the effect of using a larger value of the contagion radius, $r_{\text{c}}=10$, on the steady-state numbers obtained using Protocol A, over a range of relative relapse rates and a fixed relative transmission rate of $\omega=2$. Keeping the contagion radius fixed at $r_{\text{c}}=10$, using a lower value of $\omega$, or a different protocol results in a nearly identical plot. We note that at a large value of the contagion radius, there is essentially no distinction between the steady-state predictions of the two protocols. The long-time population of susceptibles is practically zero and the particles are split between the infected and recovered categories. This indicates that when an infected particles has an abundance of susceptible neighbors to transmit the diseases, it is immaterial if the disease spreads via the one-to-one or one-to-many route, and there are negligible susceptible individuals remaining in the long time limit. 

In the limit of a large contagion radius, each particle can ``see" all the other particles in the box, and one could therefore expect that the effect of spatial heterogeneity is reduced, bringing the microscopic model predictions closer to that obtained from the macroscopic model. Probing this line of thought, we note a qualitative similarity between Fig.~\ref{fig:rc_10_nums} and the macroscopic model results given by Fig.~\ref{fig:stead_pop}~(b), in that the steady state numbers are independent of the relative relapse rate. A distinction between the microscopic and macroscopic model predictions is that while the former predicts a vanishing of the susceptible population across the range of the relative relapse rates considered, the latter predicts a finite non-zero value for the steady-state susceptible population.

{In Fig.~\ref{fig:micro_model_alph}, the macroscopic and microscopic model results are plotted simultaneously, with the caveat that even though the same numerical values have been used for the epidemiological constants (e.g. $\alpha=\alpha^{*}=0.05$), there is no direct mapping between the two models}. Keeping the relative relapse rate $\mu$ fixed, increasing the relative transmission rate $\omega$ drives the spread of infection from a state of extinction ($S_{\infty}=1$) to one in which the fraction of susceptible individuals has reduced considerably. As noted in the discussion of Fig.~\ref{fig:micro_model}~(c), the relative transmissibility at which the transition away from the epidemic extinction state occurs depends on the value of $\mu$ in the microscopic model. For the macroscopic model, however, the location of this transition is fixed at the analytically determinable value of $\omega^{*}=1$, and is independent of the relative relapse rate $\mu$. Fig.~\ref{fig:micro_model_alph} illustrates the effect of the contagion radius on the location of this bifurcation: higher values of $r_{\text{c}}$ push the transition to lower values of $\omega$. This makes intuitive sense: a smaller infection rate is required when the infected particles can see a larger number of the susceptible population, resulting in a more effective spread of the disease. 

\begin{figure}[t]
\begin{center}
\begin{tabular}{c c}
\includegraphics[width=3in,height=!]{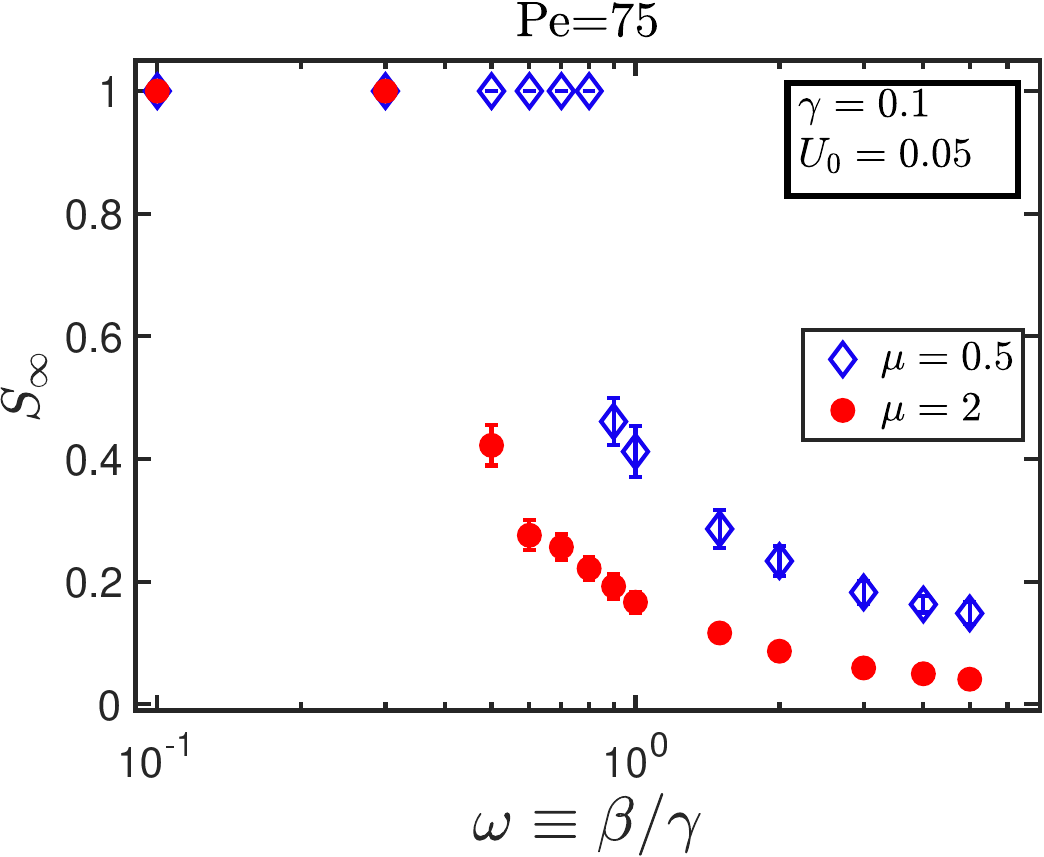}&
\includegraphics[width=3in,height=!]{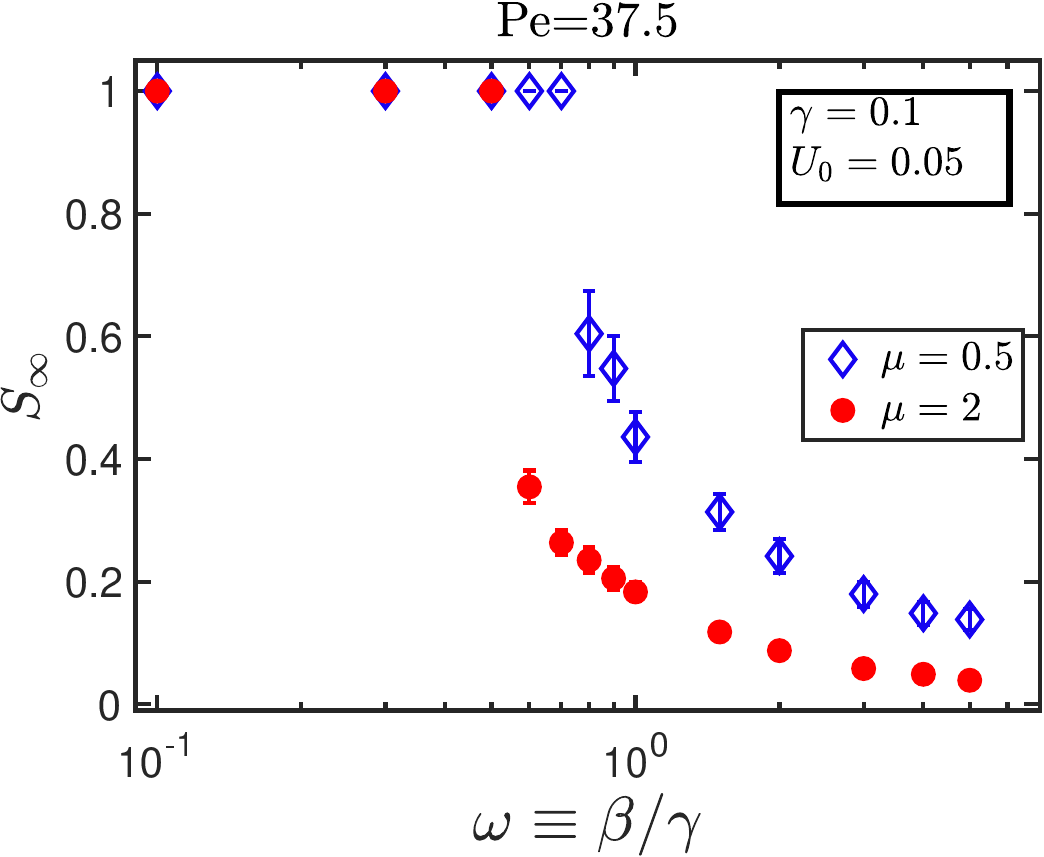}\\
(a)&(b)\\
\end{tabular}
\end{center}
\caption{Steady-state value of susceptibles as a function of the relative recovery rate, for agents moving at a slower self-propulsion speed $U_0=0.05$,  with (a) Pe=75, at which MIPS is observed, and (b) Pe=37.5, at which no MIPS is observed for $\phi_{0}=0.5$~\cite{Stenhammar2014}. The contagion radius used in both (a) and (b) is $r_{\text{cut}}=3$. Each data point in the figure was obtained from simulations of $n=5\times 10^{4}$ steps.}
\label{fig:small_v_sims}
\end{figure}

{We have examined additional factors which could determine the bifurcation point in the microscopic model. The contagion dynamics for agents moving with a reduced self-propulsion speed $U_{0}=0.05$, for two cases is analyzed. In the first case, the rotational diffusivity is set to $D_{\text{r}}=10^{-3}$ so that the P\'{e}clet number remains at $Pe=75$, at which MIPS is observed for an area fraction of $\phi_{0}=0.5$~\cite{Stenhammar2014}. In the second case, the rotational diffusivity is left unchanged at $D_{\text{r}}=2\times 10^{-3}$, so that the P\'{e}clet number falls below the threshold required for observing MIPS at $\phi_{0}=0.5$. From Fig.~\ref{fig:small_v_sims}, we see that the crossover point in all the cases is not only a function of the relapse rate and the contagion radius, but also depends on the mobility of the agent ($U_{0}$ and $D_{\text{r}}$) which indirectly determines the local packing fraction in the box.}

{The connection between the well-mixed model and the agent-based model has been examined in detail by~\citet{Paoluzzi2021}. They consider mobile agents on a two-dimensional lattice (in a periodic box of size $L$) that undergo SIR dynamics. The length of the steps is governed by the L\'{e}vy exponent (called $\beta$ in their work, but we will use the symbol $\lambda$, to avoid confusion with the rate of infection), and the step direction is chosen from a uniform random distribution. In the limit of large mobility coefficient $\lambda\to2$, the motion of the agents is akin to Brownian motion, while the $\lambda\to1$ corresponds to a L\'{e}vy flight where the agents can take steps whose lengths are picked at random from the interval $[0,L/4]$ using Mantegna's algorithm~\cite{Mantegna1994}. The lower values of $\lambda$ are seen to agree with the analytical results for the well-mixed SIR model. ~\citet{Paoluzzi2021} also studied the effect of a mixture (high and low $\lambda$) of the mobility coefficients on the contagion dynamics. They find that even a small number of sites with a small value of $\lambda$ (meaning higher mobility) can trigger epidemic waves. Note that the agents have no finite-size and hence no steric repulsion exists between them. The step sizes are entirely user-defined, and drawn from a known distribution.
In our work, although the step size is uniform \textit{by design} [with a value of $U_{0}\Delta t$ per unit time, as seen from eq.~(\ref{eq:pos_dyn}), where $U_0$ is the self-propulsion speed of the ABP], the \textit{actual} sizes of the steps vary due to steric repulsions between the agents. In any case, the maximum size of the step taken in any timestep is smaller than the particle diameter ($d=2$), and our simulation box dimensions ($L=100$) are such that $d\ll L$. These step sizes are far smaller than the ones encountered by~\citet{Paoluzzi2021}. The absence of such long steps is perhaps the reason why the microscopic model in our case does not completely converge to the well-mixed model results, although qualitative similarities are observed when the contagion radius is increased.}

{A common paradigm to study the spread of epidemic is to use a network-based approach~\cite{Kiss2017,Grossmann2021,7393962,Pastor-Satorras2015}, in which the members of the population are represented by nodes, and their connectivity denoted by the edges that join them. The infectiousness of the disease-causing vector is allowed to be different for each node in the network. This approach allows the decoupling of the connectivity of the agents from the probability of disease transmission. In the active particle-based model considered in our paper, although the input parameters for the agent mobility and the epidemiological constants are picked independently, the coupling of their effects is an $emergent$ phenomenon, due to the protocol of disease spread which depends on the spatial positioning of the various disks. ~\citet{Grossmann2021} consider SIR-dynamics on a static network where the infectiousness of the nodes can take on a distribution of values. They find that a large variation in the infectiousness leads to a smaller final size of the epidemic, stemming from an increased probability of epidemic extinction, and therefore a lowering of the herd immunity threshold~\cite{Gomes2022}. In our paper, increasing the contagion radius appears to result in a similar outcome, as evinced in Fig.~\ref{fig:micro_model_alph}, as larger $r_{\text{c}}$ results in smaller values of the transmission rate for making the infection endemic, i.e., $S_{\infty}<1$.}

We conclude this section by noting that the precise mapping (if indeed one exists) between the rate constants used in the ABM and those appearing in the ordinary differential equations at the population-level remains unknown, we predict that the contagion radius $r_{\text{c}}$ and the transmission protocol would crucially affect this relationship.

\section{\label{sec:phase_diag} Phase separation in motility-modified SIRS model}

\begin{figure*}[t]
\begin{center}
\includegraphics[width=17cm,height=!]{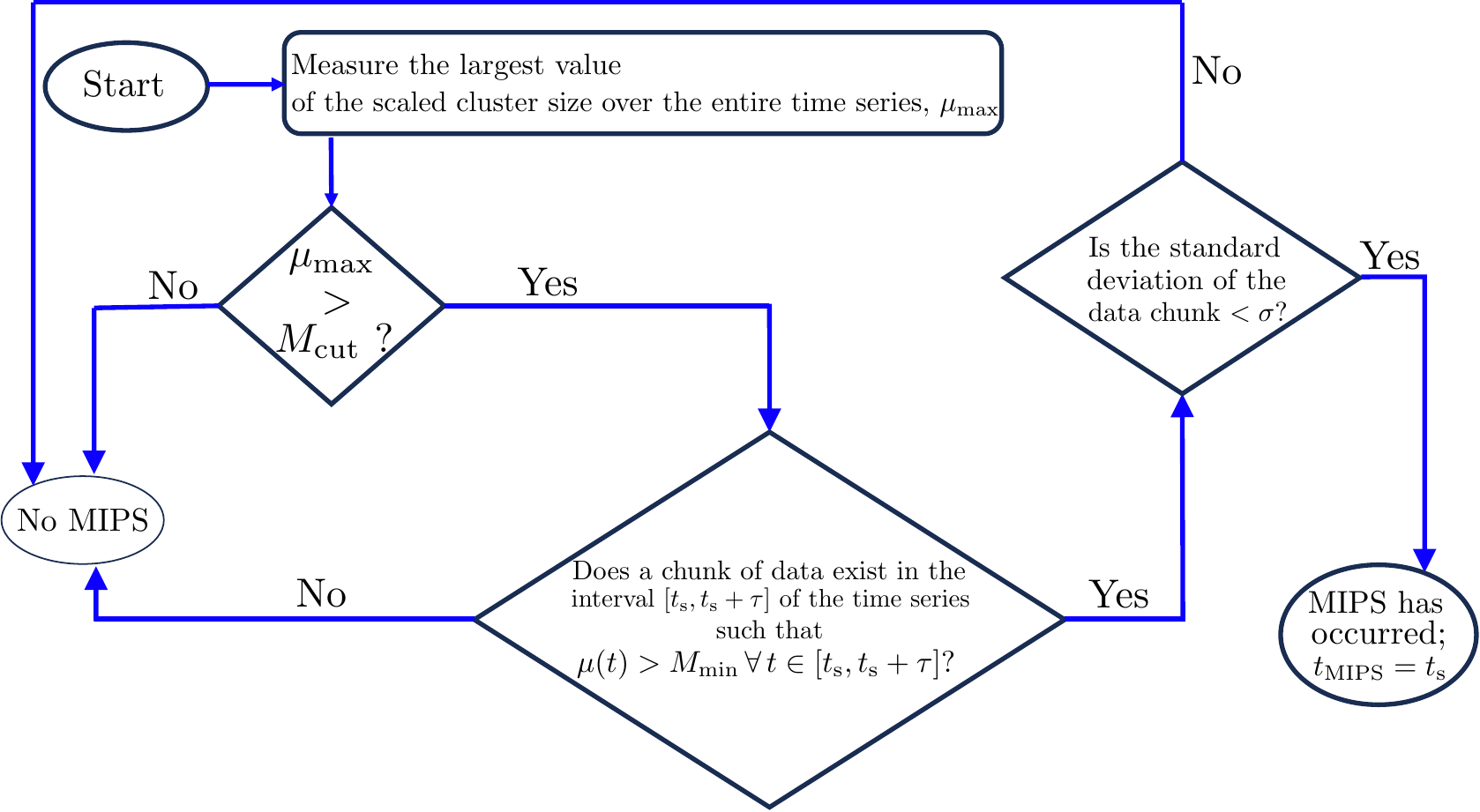}
\end{center}
\caption{Flowchart illustrating the algorithm for ascertaining if a system has undergone MIPS, and to evaluate the time needed for phase separation in case it has.}
\label{fig:mips_algo_2}
\end{figure*}

\begin{figure}
\begin{center}
\includegraphics[width=8.3cm,height=!]{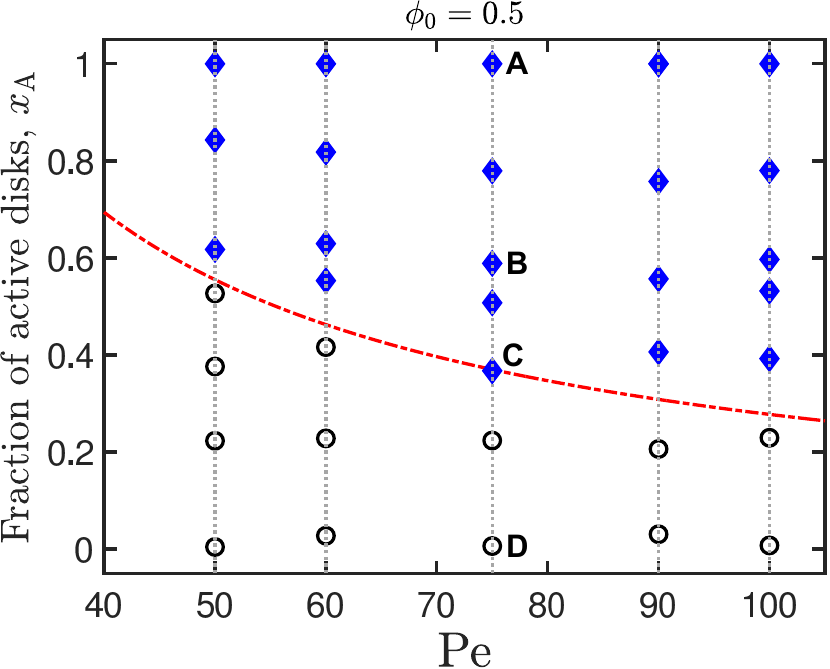}
\end{center}
\caption{Phase separation in a mixture of active and passive disks. Open circles indicate a homogeneous phase, while closed diamonds represent a phase-separated system. Dash-dotted redline represents eq.~(\ref{eq:phase_line}) with $\kappa=1.875$. Snapshots of the system at the locations A,B,C,D are given in Fig.~\ref{fig:snapshots_phase_map}.}
\label{fig:phase_map}
\end{figure}

A collection of self-propelled (or active) particles interacting sterically undergo a motility induced phase separation (MIPS) at large enough values of the P\'{e}clet number and the area fraction $\phi_{0}$ of the particles. In this transition, the particles go from being in a gas-like, single-phase to a phase-separated state consisting of a dense large cluster that dynamically exchanges particles with the surrounding dilute phase [see refs.~\citenum{Cates2015,Zottl2023} for an extensive review of the topic]. The boundary separating the homogeneous state from the phase-separated one in the Pe-$\phi_0$ plane has been determined through direct numerical simulations~\cite{Redner2013,Stenhammar2014,Levis2017} and analytical theory~\cite{Stenhammar2013,Redner2013,Bialke2013,Takatori2015_theo,Worlitzer2021}.

The effect of the presence of passive particles - ones that move translationally under the effect of Brownian noise or not at all - on the phase separation behavior of an active-passive mixture has also received interest~\cite{Stenhammar2015,Takatori2015,RogelRodriguez2020,Agrawal2021,Zhang2022}. ~\citet{Stenhammar2015} studied such a mixture with a total area fraction of $\phi_{0}$, of which a number fraction $x_{A}$ is active. They derive the following analytical expression for the phase boundary in the $\text{Pe}-x_{A}$ plane:  
\begin{equation}\label{eq:phase_line}
x_{\text{A}}=\dfrac{3\pi^2\kappa}{4\phi_{0}\text{Pe}}
\end{equation}
using the kinetic model introduced by~\citet{Redner2013}. Here $\kappa$ is an empirical fitting parameter~\cite{Redner2013} that represents the average total number of particles that are lost from a phase-separated cluster in an escape event. The boundary obtained using $\kappa=4.05$ is seen to accurately demarcate the homogeneous and demixed states in the phase diagram generated from numerical simulations at $\phi_{0}=0.6$. A value of $\kappa=4.5$ accurately predicts the phase boundary in the Pe-$\phi_{0}$ plane for a system composed solely of active disks ($x_{A}=1$). ~\citet{Takatori2015} derive an expression for the phase boundary in an active-passive mixture using an alternative approach that relies on the concept of active swim pressure in a collection of self-propelled swimmers. They obtain an agreement with the predictions of~\citet{Stenhammar2015} without any fitting parameters.

\begin{table}[t]
\setlength{\tabcolsep}{10pt}
\centering
\caption{\label{param_phase_diag} Data points in the $\text{Pe}-x_{A}$ phase plane (Fig.~\ref{fig:phase_map}) and the epidemiological constants used in the simulations for obtaining them.}
\vspace{5pt}
\begin{tabular}{ | c | c | c | c | c |}
\hline
Pe & $x_A$ & $\beta$ & $\gamma$ & $\alpha$  \\ \hline
 & $1.0$ & 0 & 0 & 0 \\ \cline{2-5}
 & $0.84$ & 0.1 & 0.1 & 0.05 \\ \cline{2-5}
 & $0.62$ & 0.05 & 0.1 & 0.2 \\ \cline{2-5}
 50 & $0.53$ & 0.2 & 0.15 & 0.2 \\ \cline{2-5}
 & $0.38$ & 0.2 & 0.1 & 0.2 \\ \cline{2-5}
 & $0.22$ & 0.2 & 0.05 & 0.2 \\ \cline{2-5}
 & $0.004$ & 0.2 & 0.001 & 0.2 \\ \cline{2-5} 
 \hline
  & $1.0$ & 0.05 & 0.1 & 0.05 \\ \cline{2-5}
 & $0.82$ & 0.1 & 0.1 & 0.05 \\ \cline{2-5}
 & $0.63$ & 0.2 & 0.2 & 0.2 \\ \cline{2-5}
 60 & $0.55$ & 0.2 & 0.15 & 0.2 \\ \cline{2-5}
 & $0.42$ & 0.2 & 0.1 & 0.2 \\ \cline{2-5}
 & $0.23$ & 0.2 & 0.05 & 0.2 \\ \cline{2-5}
 & $0.03$ & 0.2 & 0.005 & 0.2 \\ \cline{2-5} 
 \hline
 & $1.0$ & 0 & 0 & 0 \\ \cline{2-5}
 & $0.78$ & 0.1 & 0.1 & 0.05 \\ \cline{2-5}
 & $0.59$ & 0.2 & 0.2 & 0.2 \\ \cline{2-5}
 75 & $0.51$ & 0.05 & 0.1 & 0.2 \\ \cline{2-5}
 & $0.37$ & 0.2 & 0.1 & 0.2 \\ \cline{2-5}
 & $0.22$ & 0.2 & 0.05 & 0.2 \\ \cline{2-5}
 & $0.006$ & 0.2 & 0.001 & 0.2 \\ \cline{2-5} 
 \hline
   & $1.0$ & 0.05 & 0.1 & 0.05 \\ \cline{2-5}
 & $0.76$ & 0.1 & 0.1 & 0.05 \\ \cline{2-5}
 & $0.56$ & 0.2 & 0.15 & 0.2 \\ \cline{2-5}
 90 & $0.41$ & 0.2 & 0.1 & 0.2 \\ \cline{2-5}
 & $0.21$ & 0.2 & 0.05 & 0.2 \\ \cline{2-5}
 & $0.03$ & 0.2 & 0.005 & 0.2 \\ \cline{2-5}
 \hline
   & $1.0$ & 0 & 0 & 0 \\ \cline{2-5}
 & $0.78$ & 0.1 & 0.1 & 0.05 \\ \cline{2-5}
 & $0.59$ & 0.2 & 0.2 & 0.2 \\ \cline{2-5}
 100 & $0.53$ & 0.05 & 0.1 & 0.2 \\ \cline{2-5}
 & $0.39$ & 0.2 & 0.1 & 0.2 \\ \cline{2-5}
 & $0.23$ & 0.2 & 0.05 & 0.2 \\ \cline{2-5}
 & $0.008$ & 0.2 & 0.001 & 0.2 \\ \cline{2-5} 
 \hline
\end{tabular}
\end{table}

\begin{figure}[t]
\begin{center}
\includegraphics[width=3.5in,height=!]{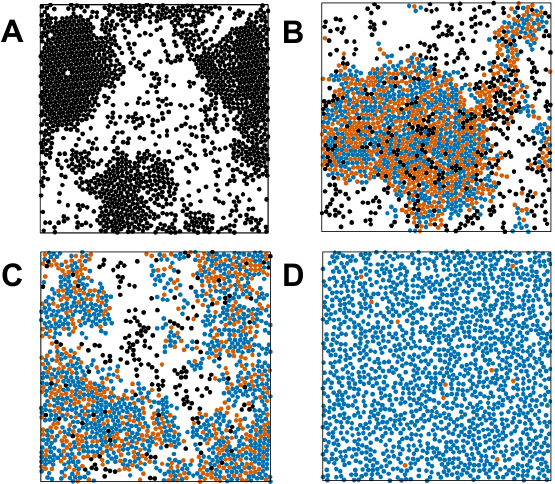}
\end{center}
\caption{Representative snapshots of the system for $\text{Pe}=75$ at the active-disk fraction (A-D) indicated in Fig.~\ref{fig:phase_map} and recorded at the halfway point of the total simulation. The black disks denote susceptible agents, while blue and orange disks represent infected and recovered individuals, respectively.}
\label{fig:snapshots_phase_map}
\end{figure}

\begin{figure}[b]
\begin{center}
\includegraphics[width=3.5in,height=!]{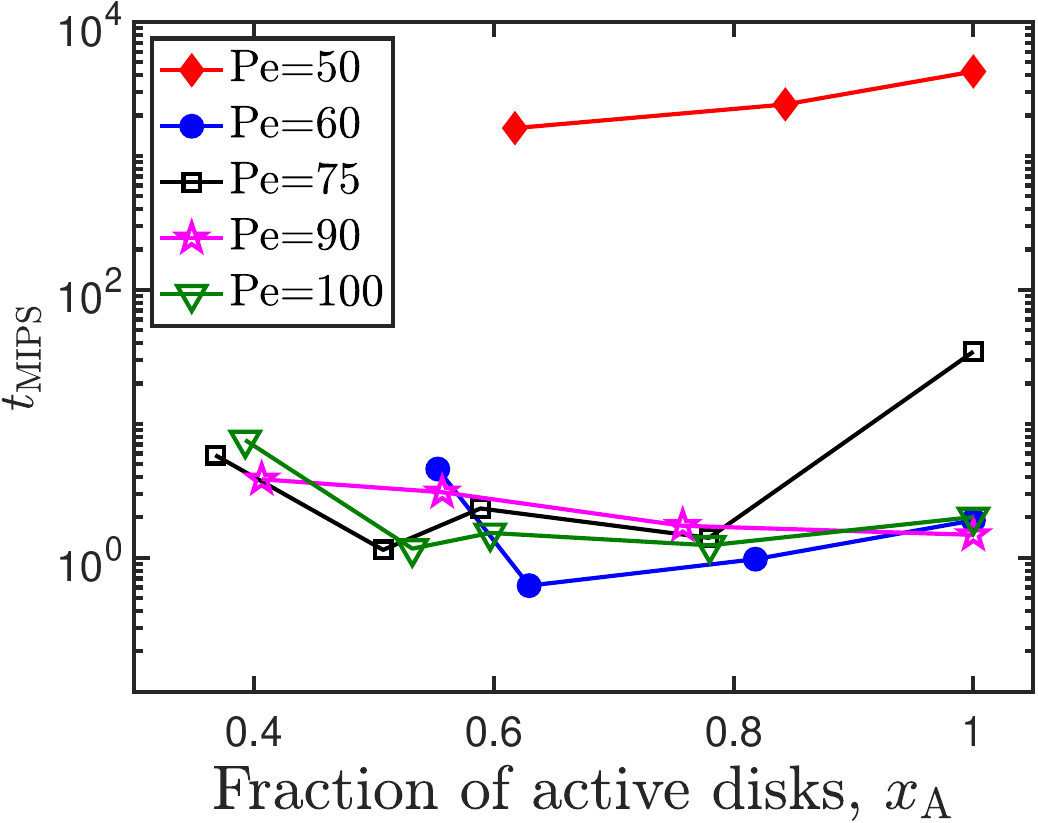}
\end{center}
\caption{Time taken for motility induced phase separation, as a function of the active disk fraction.}
\label{fig:t_mips}
\end{figure}

\begin{figure}[t]
\begin{center}
\begin{tabular}{c c}
\includegraphics[width=3in,height=!]{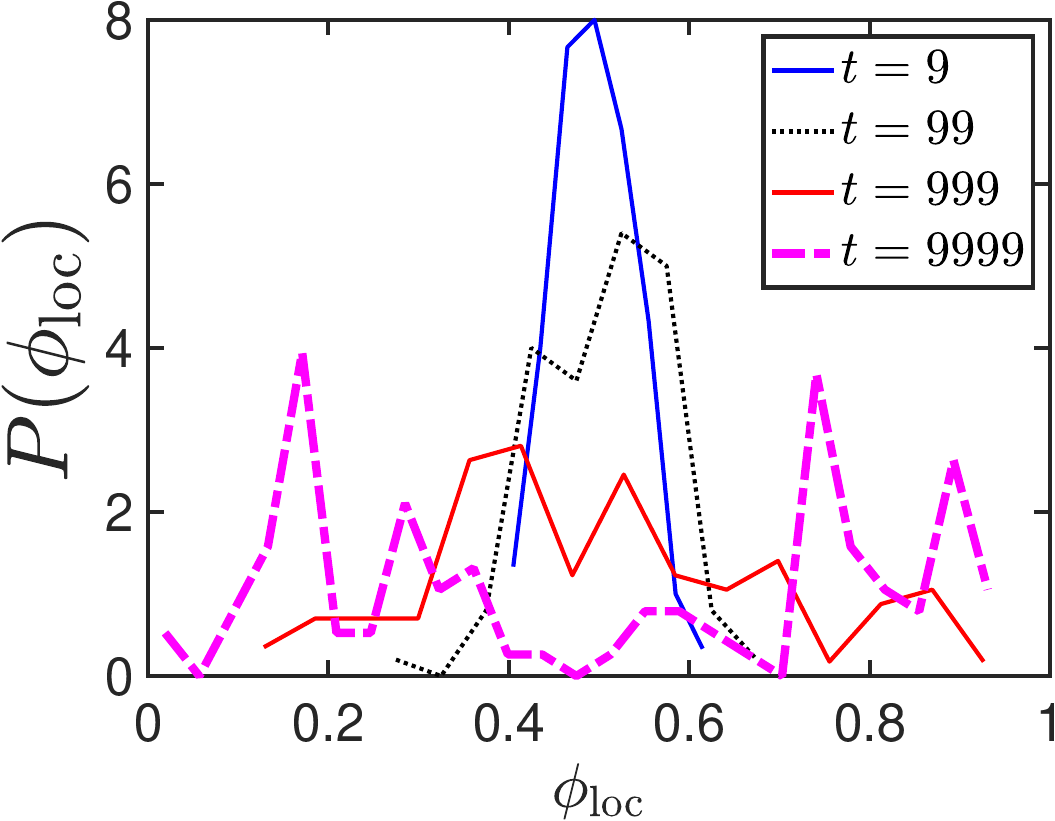}&
\includegraphics[width=3in,height=!]{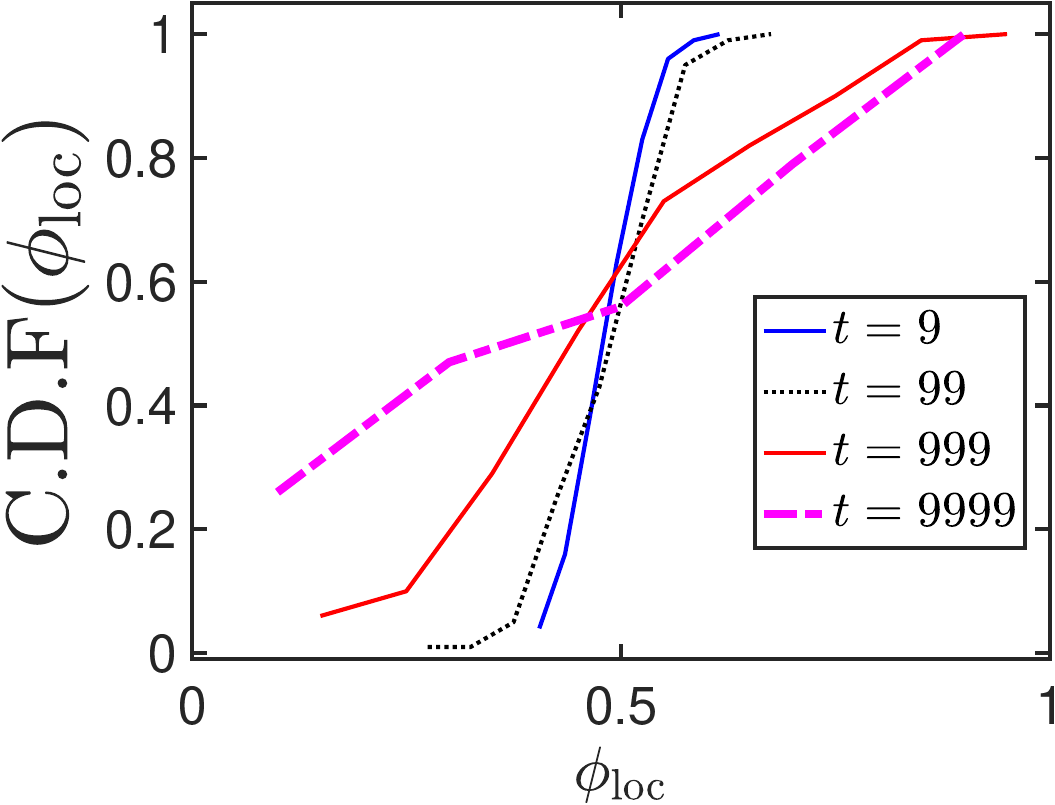}\\
(a) & (b)\\[5pt]
\end{tabular}
\end{center}
\caption{(a) Probability and (b) cumulative distribution functions of the local area fraction, for the motility induced phase separation process represented in Fig.~\ref{fig:base_case_mips}.}
\label{fig:mips_dist_func}
\end{figure}

\begin{figure}[t]
\begin{center}
\begin{tabular}{c c}
\includegraphics[width=3in,height=!]{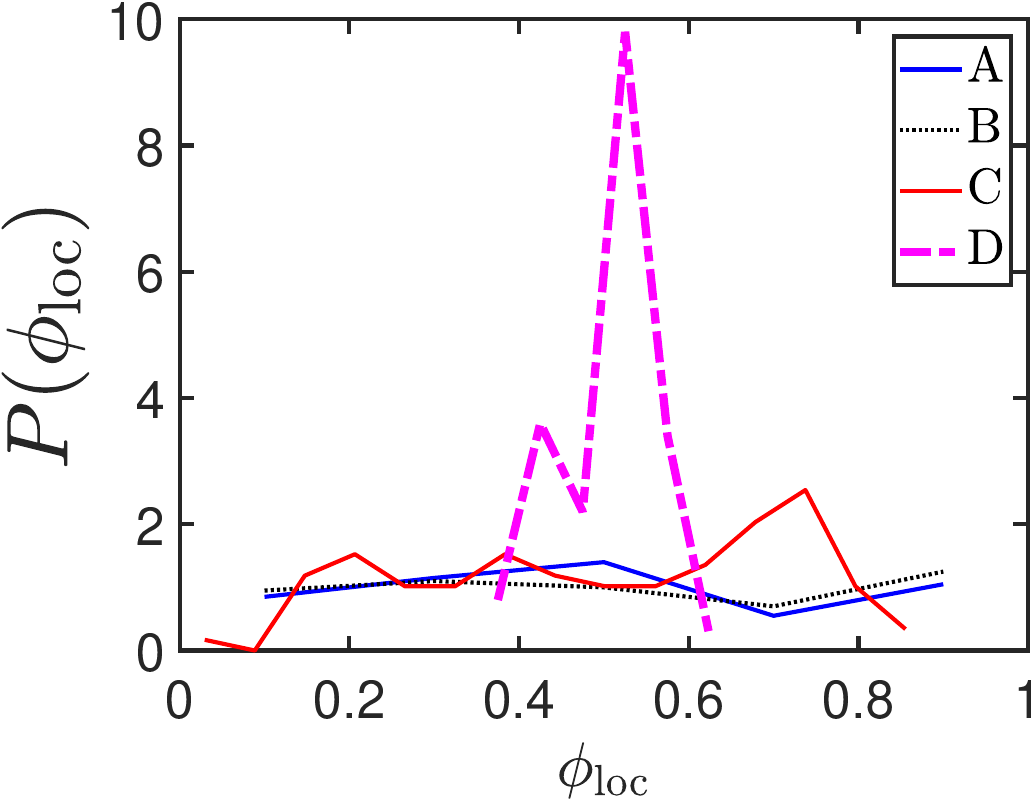}&
\includegraphics[width=3in,height=!]{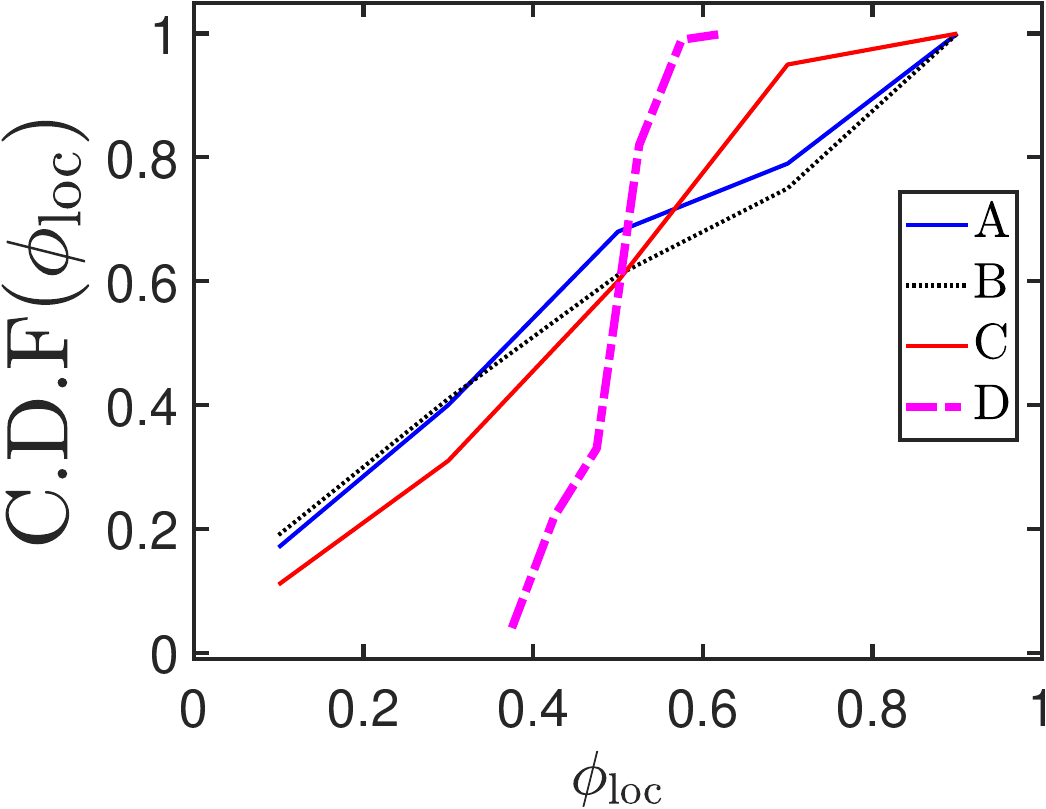}\\
(a) & (b)\\
\end{tabular}
\end{center}
\caption{(a) Probability and (b) cumulative distribution functions of the local area fraction, for the snapshots given in Fig.~\ref{fig:snapshots_phase_map}.}
\label{fig:abcd_dist_func}
\end{figure}

\begin{figure}[b]
\begin{center}
\includegraphics[width=3.5in,height=!]{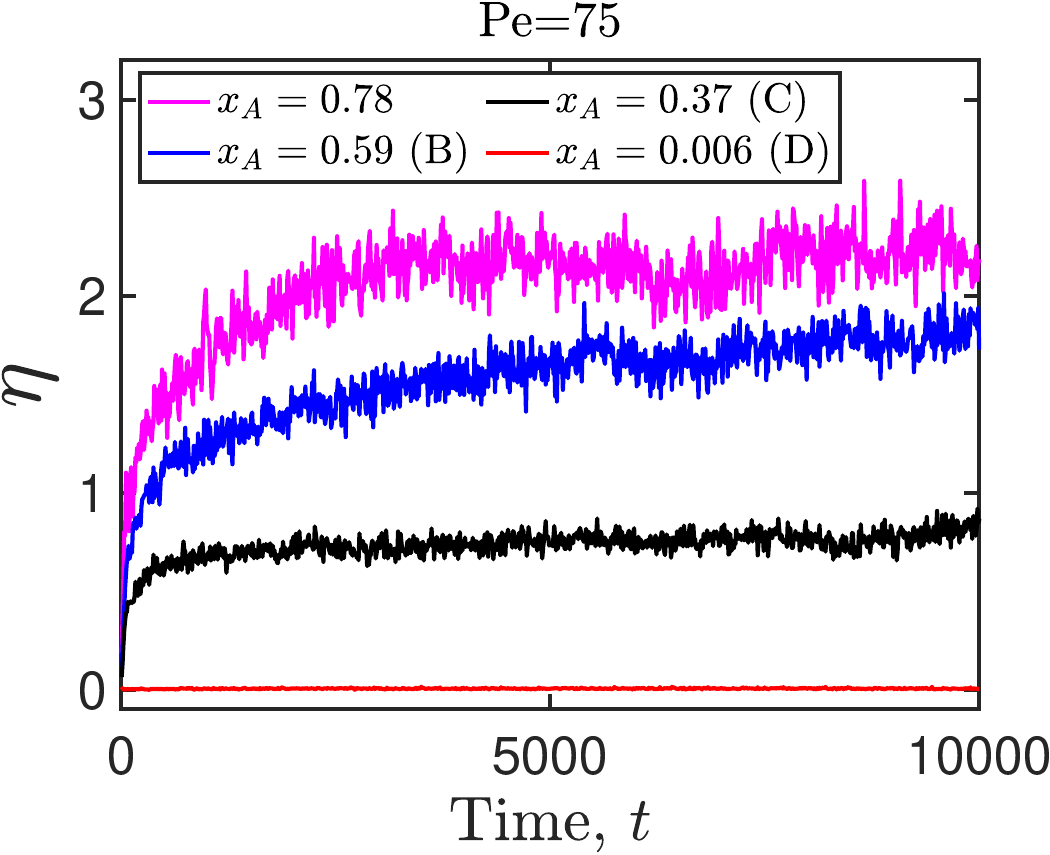}
\end{center}
\caption{Average nearest neighbors $\eta$ as a function of time for Pe=75 at various values of the steady-state active fractions. The alphabets in parentheses in the legend entry correspond to the points indicated in Fig.~\ref{fig:phase_map}.}
\label{fig:eta_pe_75}
\end{figure}

\begin{figure*}[t]
\begin{center}
\begin{tabular}{c c}
\includegraphics[width=3in,height=!]{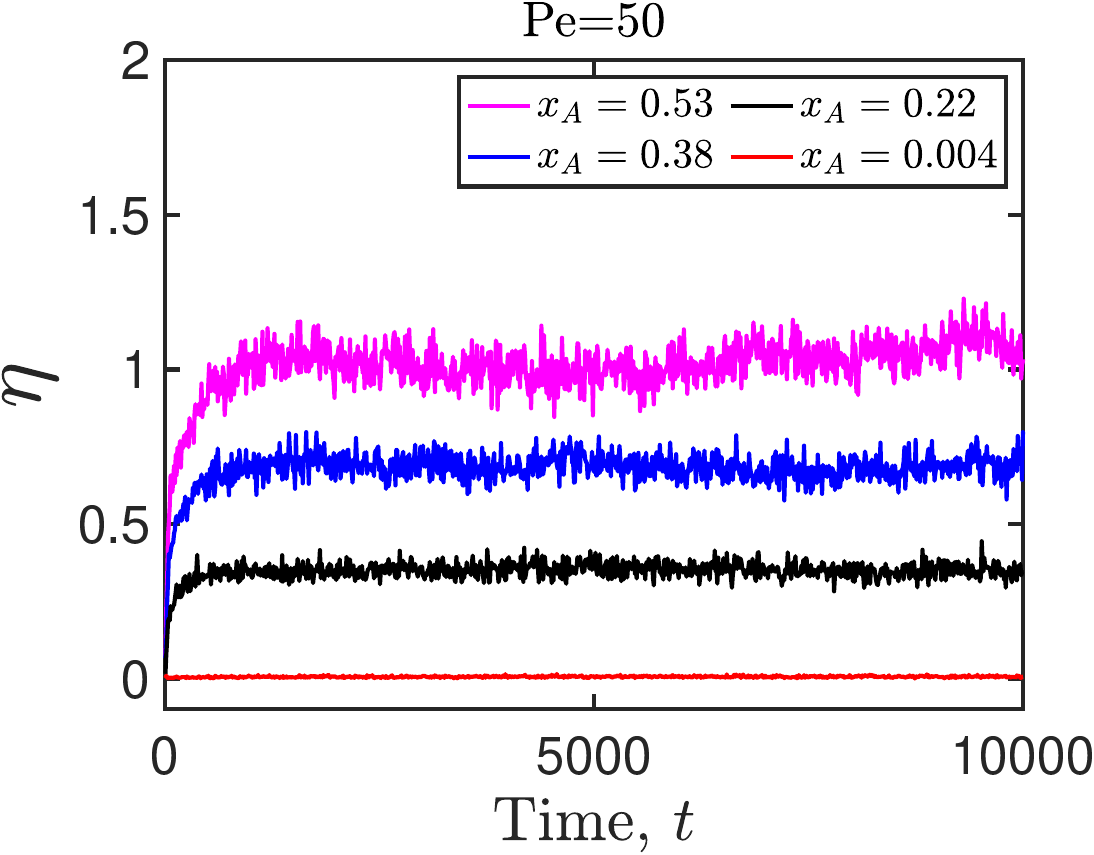}&
\includegraphics[width=3in,height=!]{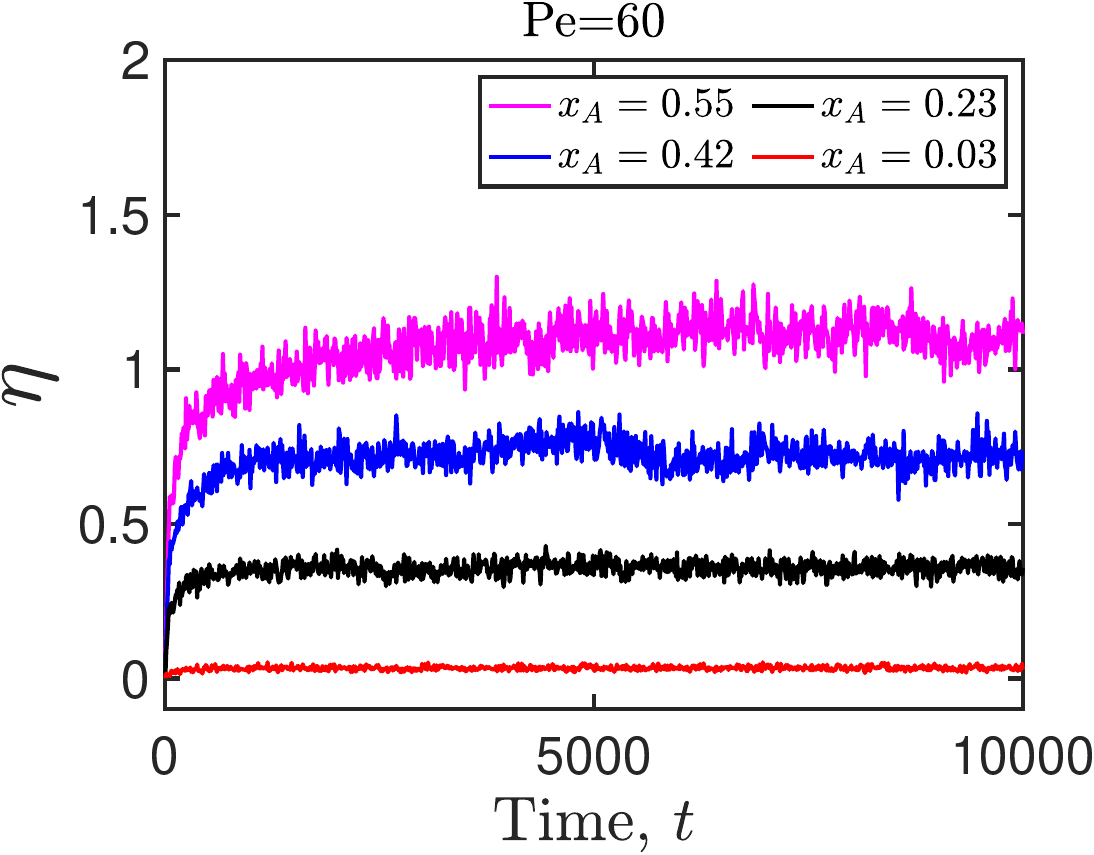}\\[5pt]
(a) & (b)\\ [10pt]
\includegraphics[width=3in,height=!]{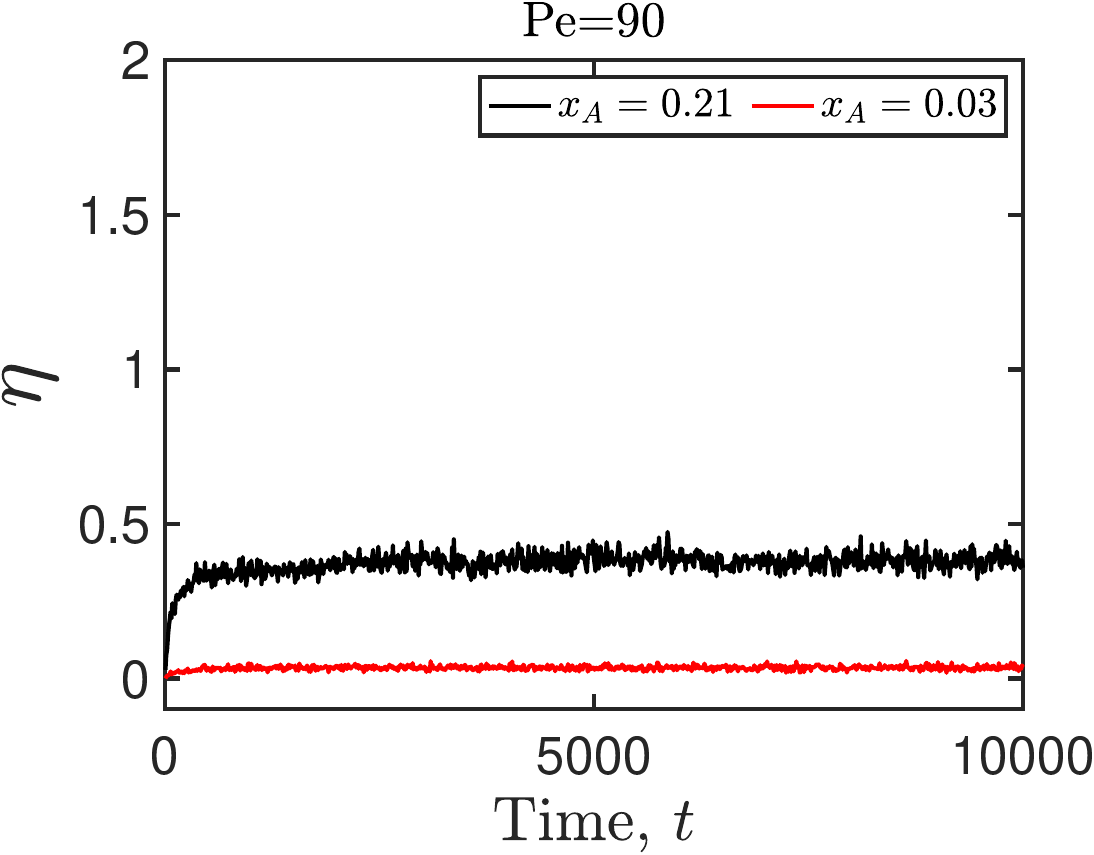}&
\includegraphics[width=3in,height=!]{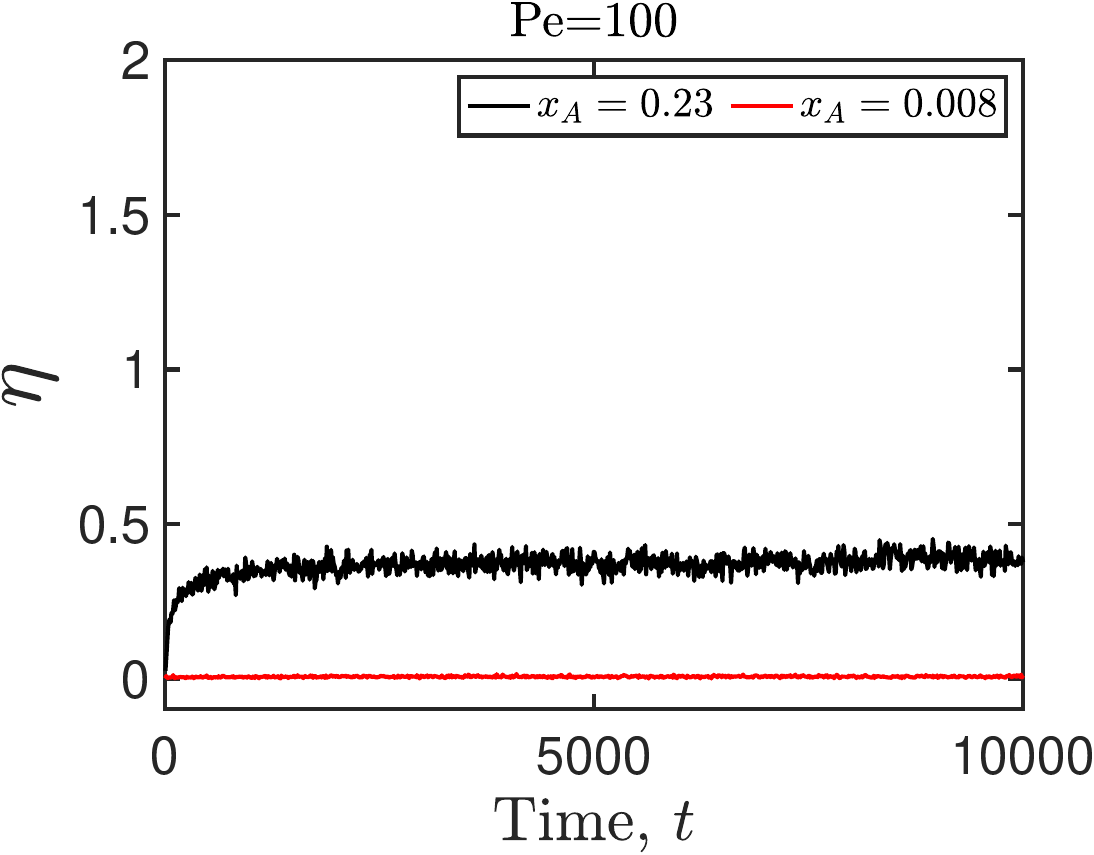}\\[5pt]
(c) & (d) 
\end{tabular}
\end{center}
\caption{Average nearest neighbors $\eta$ as a function of time for (a) Pe=50, (b) Pe=60, (c) Pe=90, and (d) Pe=100, at various number fractions of the active disks ($x_A$).}
\label{fig:eta_diff_pe}
\end{figure*}

We summarize the common metrics used in the literature to quantify motility-induced phase separation, before describing the algorithm introduced in the present work to identify systems that have undergone phase separation. The fraction of particles in the largest cluster $N_{\text{LC}}/N$ is a popular metric~\cite{Patch2017,Su2023,Kailasham2023_coll} to track the approach to MIPS. When the majority of the particles in the system belong to a single cluster, it is taken to be an indication of MIPS. Large fluctuations in the number of particles within a subregion are taken to be a sign of inhomogeneity and the onset of MIPS~\cite{Fily2012,Desgranges2023}. Another signature for the occurrence of MIPS is the appearance of a bimodality in the probability distribution of the local area fraction~\cite{Gonnella2015,Knezevic2022,Theeyancheri2024}. As a system evolves from a homogeneous state to a phase separated one, the size of the domains (calculated from either the static structure factor or the density correlation function) grows with time as a power law~\cite{Redner2013,Stenhammar2014,Caporusso2022,Caporusso2020} $\mathcal{L}(t)\sim t^{1/3}$.
These metrics have also been used to analyze systems containing a mixture of active and passive particles~\cite{Stenhammar2015}, in which the relative populations of the two species remain \textit{constant} in time. 

In the present work, however, the fraction of active and passive disks $fluctuate$ in time due to the disease spreading, and we seek to identify an appropriate metric for the identification of MIPS in such systems. To that end, Fig.~\ref{fig:base_case_mips} illustrates the time evolution of $N_{\text{LC}}/N$ in a system consisting entirely of susceptible particles, with the epidemiological constants set to zero. There is no spread of infection in such a system, and the fraction of active disks is therefore unity at all times. For the values of $\text{Pe}$ and the area fraction considered in Fig.~\ref{fig:base_case_mips}, the system undergoes MIPS, as evinced by the snapshots recorded at the various time instances [Fig.~\ref{fig:snapshots_base_case_mips}]. Our goal is to define metrics for the identification of MIPS based on this reference time series, for application to other systems in our work in which the fraction of active disks fluctuate in time. Firstly, we note that the normalized size of the largest cluster reaches a steady state value of $N_{\text{LC}}/N\approx0.83$ following an initial transient. Secondly, after $t\approx 5520$, there are no significant dips in the value of $N_{\text{LC}}/N$, and the fluctuations in this quantity are minimal. We use these two observations to devise a methodology (Fig.~\ref{fig:mips_algo_1} and Fig.~\ref{fig:mips_algo_2}) for ascertaining if a system has undergone MIPS or not, given the time-series of the largest cluster. In case a system has phase separated, this algorithm also estimates the time needed for MIPS. The various parameters needed by the algorithm are $\{M_{\text{cut}},M_{\text{min}},\tau,\sigma\}$, and a brief explanation is as follows. If the fractional size of the largest cluster $\mu\equiv N_{\text{LC}}/N$ does not exceed $M_{\text{cut}}$ at any point in its time series, then we consider that MIPS has not occurred.
The algorithm searches for a chunk of data in the time series in the interval $\left[t_{\text{s}},t_{\text{s}}+\tau\right]$, such that each data point in the interval exceeds $M_{\text{min}}$. If such a chunk is not found in the input data series, the algorithm concludes that MIPS has not occurred. Provided such a data chunk is found, we then test if the standard deviation of the data series (normalized by the total number of particles), is smaller than $\sigma$. If this requirement is met, the algorithm concludes that MIPS has occurred and returns $t_{\text{s}}$, the starting point of the data chunk, as the time at which MIPS has occurred. If the standard deviation of the data series (normalized by the total number of particles) exceeds $\sigma$, then we conclude that MIPS has not occurred. The parameter values used in this algorithm are listed in Table~\ref{param_MIPS}. The time to MIPS as estimated by the algorithm for the timeseries indicated in Figs.~\ref{fig:base_case_mips} and ~\ref{fig:mips_algo_1} are 3730 and 5527, respectively.

Fig.~\ref{fig:phase_map} explores the phase behavior of the system in the $\text{Pe}-x_{A}$ plane, obtained from simulations using a wide range of the epidemiological constants (see Table~\ref{param_phase_diag} for values). At the steady-state value of the fraction of active disks for any given value of the P\'{e}clet number, we denote if MIPS has occurred or not using the algorithm described above. The boundary between the homogeneous states and the MIPS states is well described by eq.~\ref{eq:phase_line} when a value of $\kappa=1.875$ is used. Snapshots of the system at a fixed value of $\text{Pe}$ and varying fractions of the active disks are given in Fig.~\ref{fig:snapshots_phase_map}~A-D. We notice that there is no preference for disks with identical internal states to cluster together. A more quantitative analysis would involve the calculation of the pair correlation function for the various populations.

In Fig.~\ref{fig:t_mips}, the time taken for motility induced phase separation, as identified using the algorithm described in Fig.~\ref{fig:mips_algo_2}, is plotted as a function of the fraction of active disks in the system, for a range of P\'{e}clet numbers. We observed no definitive trend, implying that we cannot conclude if the presence of transiently immobile disks helps to aid or suppress phase separation in an active-passive mixture. ~\citet{Forgacs2022} observe that the presence of quenched disorder, or immobile obstacles, in an active matter system causes the formation of numerous small clusters in addition to the large cluster that is characteristic of MIPS. Additionally, MIPS is a re-entrant phenomenon~\cite{Bialke2013,Su2023}, meaning that increasing the P\'{e}clet number in an already phase separated system can cause the system to go back to being in a homogeneous phase.

{Another common metric to track motility induced phase separation is the local area fraction $\phi_{\text{loc}}$. This is evaluated by dividing the periodic box into multiple smaller boxes and measuring the area fraction occupied by disks in each sub-box, to get a distribution of $\phi_{\text{loc}}$ values~\cite{Theeyancheri2024}. In Fig.~\ref{fig:mips_dist_func} and ~\ref{fig:abcd_dist_func}, the probability and cumulative distribution functions of the local packing fraction (respectively), for the data series represented in Fig.~\ref{fig:base_case_mips} of the manuscript, at various timepoints. The onset of MIPS is indicated by the appearance of bimodality in the probability distribution function. The CDF peaks sharply around the average packing fraction ($\phi_0=0.5$) in the absence of MIPS, and is seen to broaden as MIPS progresses (Fig.~\ref{fig:abcd_dist_func}).}

{We have also calculated the average number of mobile (susceptible and recovered) agents that are nearest neighbors to a an immobile (infected) agent at a given time ($t$), similar to the methodology adopted by~\citet{Forgacs2022}. This metric is calculated as follows: 
\begin{equation}\label{eq:eta_def}
\eta(t)=\dfrac{1}{I(t)}\sum_{i}^{I(t)}\sum_{j}^{S(t)+R(t)}\mathbb{I}\left(|r_{ij}(t)|=d\right)
\end{equation}
where the indicator function $\mathbb{I}(\cdots)$ returns 1 (0) if its argument is true (false). The numerical implementation of eq.~(\ref{eq:eta_def}) allows for a $0.5\%$ tolerance, and uses a value of $1.005d$ in place of $d$. In Fig.~\ref{fig:eta_pe_75}, the average number of neighbors is evaluated for the Pe=75 case, for steady-state active fractions ($x_A$) on either side of the phase boundary. We note that point A in the phase diagram is obtained for a system in which all the particles are of the susceptible type, with the epidemiological constants set to zero. There are no infected particles in this case, and $\eta(t)$ is therefore not defined. It is clear from the figure that the average number of neighbors drops with the decrease in the fraction of active disks in the system. We performed a similar analysis for all the other P\'{e}clet numbers we have examined in the current study, and focused on those points at which MIPS has not occurred. It is clear that the infected particles precipitate the formation of microclusters, even in the absence of a global motility induced phase separation (Fig.~\ref{fig:eta_diff_pe}).}

\section{\label{sec:concl} Conclusions}

We performed an agent-based modeling of disease spread according to the SIRS model using a collection of active Brownian particles moving in two dimensions whose internal state encodes their state of infection. Two protocols for infection were considered, and their efficacies for the spread of the disease were analyzed for various combinations of the epidemiological constants. The coupling of the particle's internal state to its motility causes the population to behave as a collection of particles in which the fraction of active disks is time-dependent. We developed an algorithm to determine the occurrence of motility induced phase separation in such systems with \textit{transient} activity, and find that it is well-described by the theories for phase separation in active-passive mixtures where the fraction of active disks remains constant in time. Although a direct mapping between the agent-based (microscopic) and macroscopic model is not found, several common features between the contagion dynamics predicted by the two models are noted. We see evidence for a transcritical bifurcation in the microscopic model where the agents are modeled as active Brownian particles. The use of active Brownian disks permits a tractable method to tune the density distribution of the system by changing the P\'{e}clet number. Humans in general, however, do not move at a constant self-propulsion speed with randomly varying orientations. Simulating the dynamics of individuals in a crowd has typically relied on the use of social forces that describe the interaction between the individual members~\cite{Teknomo2006,Miller2007}. The use of such pedestrian models to describe the motion of individual agents could permit the extension of the present work to model epidemic spread in human populations. Another interesting exercise for future work could be the effect of the type of motility on the nature and location of the bifurcation point, i.e., would a system of agents modeled using social forces exhibit a different kind of bifurcation when the steady-state numbers are plotted as a function of the relative rate of transmission. {The concept of over-dispersion has been observed in the case of the COVID-19 pandemic~\cite{Grossmann2021}, in which a few members of the population transmit the infection to many, while most individuals infect only a few or none at all. This aspect could be included in our framework by prescribing that certain infected agents in the system to follow Protocol A (one-to-one), while a few others follow Protocol B (one-to-many).} 

{\section*{Author contributions}}
{R.K. and A.K designed research, performed research, contributed new analytic tools, analyzed data, and wrote the paper.}

\section*{Conflicts of interest}
There are no conflicts to declare.

\section*{Data and Code availability}
MATLAB codes for the solution of the macroscopic epidemiological model and HOOMD-blue~\cite{Anderson2020} scripts for agent-based modeling are available freely on GitHub~\cite{Kailasham_PM_and_RS_2024}. Visualization of simulation data was performed using OVITO~\cite{Ovito2010}. Simulation data are available upon request from the authors.

\section*{Acknowledgements}
We gratefully acknowledge the support of the Charles E. Kaufmann Foundation of the Pittsburgh Foundation (Grant No. 1031373-438639). R. K. thanks Ligesh Theeyancheri for valuable insights regarding the calculation of the local area fraction.

\bibliography{ms_v2} 
\bibliographystyle{rsc} 

\end{document}